# ABI Approach: Automatic Bias Identification in Decision-Making Under Risk based in an Ontology of Behavioral Economics

Eduardo da C. Ramos
Graduate Program in Informatics, Universidade Federal do Rio de Janeiro, Brazil, eramos@ufrj.br

Maria Luiza M. Campos
Graduate Program in Informatics, Universidade Federal do Rio de Janeiro, Brazil, mluiza@ufrj.br

Fernanda Baião
Department of Industrial Engineering, Pontifical Catholic University of Rio de Janeiro, Brazil, fbaiao@puc-rio.br

Organizational decision-making is crucial for success, yet cognitive biases can significantly affect risk preferences, leading to suboptimal outcomes. Risk seeking preferences for losses, driven by biases such as loss aversion, pose challenges and can result in severe negative consequences, including financial losses. This research introduces the ABI approach, a novel solution designed to support organizational decision-makers by automatically identifying and explaining risk seeking preferences during decision-making. This research makes a novel contribution by automating the identification and explanation of risk seeking preferences using Cumulative Prospect theory (CPT) from Behavioral Economics. The ABI approach transforms theoretical insights into actionable, real-time guidance, making them accessible to a broader range of organizations and decision-makers without requiring specialized personnel. By contextualizing CPT concepts into business language, the approach facilitates widespread adoption and enhances decision-making processes with deep behavioral insights. Our systematic literature review identified significant gaps in existing methods, especially the lack of automated solutions with a concrete mechanism for automatically identifying risk seeking preferences, and the absence of formal knowledge representation, such as ontologies, for identifying and explaining the risk preferences. The ABI Approach addresses these gaps, offering a significant contribution to decision-making research and practice. Furthermore, it enables automatic collection of historical decision data with risk preferences, providing valuable insights for enhancing strategic management and long-term organizational performance. An experiment provided preliminary evidence on its effectiveness in helping decision-makers recognize their risk seeking preferences during decision-making in the loss domain.

**Additional Keywords and Phrases:** Organizational decision-making under risk, Ontologies, Behavioral economics, Cumulative Prospect theory, cognitive bias, risk seeking preference

## 1 INTRODUCTION

Organizational decision-making is critical for business success, occurring in complex environments with multiple tasks (SIMON, 1997). Decision-makers often rely on intuitive decision-making, especially in uncertain situations (RIEL & HORVÁRTH, 2014). Intuition, characterized by affectively-charged judgments arising from rapid and non-conscious associations, helps tackle complex problems in dynamic and information-limited environments (DANE and PRATT, 2009). However, cognitive biases can undermine intuitive decisions, causing systematic deviations from rationality (KAHNEMAN and TVERSKY, 1972; SAMSON, 2021), as exemplified by the case of Kodak's bankruptcy, where cognitive biases played a role in its downfall (DOMEIER and SACHSE, 2015).

Commonly, decision-makers are unaware of their biases and how they influence risk preferences. Risk seeking preferences for losses of medium and high probabilities can have significant negative financial impacts on organizations. Loss aversion bias, where individuals are more averse to losses than motivated by equivalent gains, contributes to this preference. So, in loss domain, decision-makers may take risks to avoid losses, even if these decisions are irrational or disadvantageous in the long run (KAHNEMAN et al., 2021; KAHNEMAN, 2011; LUCA and BAZERMAN, 2021).

Additionally, even when decision-makers have previous knowledge of biases, it remains challenging for them to identify their own biases during the decision-making process (KAHNEMAN, 2011; RUGGERI et al., 2020). So, the problem considered in our research is the difficulty of an organizational decision-maker to identify during decision-making that they are influenced by risk seeking preference that leads them to reject the favorable alternative and to choose the riskier one.

While various approaches exist for identifying and mitigating biases (TSIPURSKY, 2019; ABATECOLA et al., 2018; CRISTOFARO, 2017; BAZERMAN and MOORE, 2013; ARNOTT, 2006), there are few approaches supporting organizational decision-makers during the decision-making process to recognize their own risk seeking preferences for losses of medium and high probabilities (KAHNEMAN et al., 2021; KAHNEMAN et al., 2011b). Moreover, the precise characterization and systematic identification of the biased risk seeking preferences are still a challenge in organizational contexts. This brings us to the main research question (RQ) of this study: "how to support an organizational decision-maker to identify that they are subjected to individual risk seeking preference during a decision-making?".

The "bias observation checklist" approach (KAHNEMAN et al., 2021; KAHNEMAN et al., 2011b) is an existing solution and is the most promising technique according to Kahneman et al. (2021). It consists of having an observer to identify biases during decision-making. However, this method relies heavily on human observers who require training to identify biases and who can be susceptible to their own cognitive biases, introducing potential errors into the process. Additionally, it may not be feasible for some organizations training or hiring a decision observer. Given these limitations, it remains a significant challenge to effectively identify risk seeking preferences during organizational decision-making.

A systematic literature review (SLR) was conducted to explore key research studies on identifying risk seeking preferences in organizational decision-making. The SLR found that while some studies addressed the identification of risk seeking preferences during decision-making, none offered a concrete mechanism for automatically identifying these risk preferences. Moreover, apart from Ohlert and Weißenberger (2020), most studies did not provide any explanation of these risk preferences to decision-makers, revealing a substantial gap in the literature on this subject. In addition, we highlight a lack of approaches using a formal representation of the knowledge related to the risk seeking preference, such as an ontology.

This combined evidence highlights a critical need for an automated solution that not only identifies but also explains risk seeking preferences during decision-making (in real-time). To address this gap and answer the main research question, we propose the ABI approach, which automates the identification and contextual explanation of risk seeking preferences during decision-making according to the Cumulative Prospect theory (CPT), which is the main theory of Behavioral Economics.

To develop the ABI approach, we created the intuitive decision-making ontology (RAMOS et al., 2021), which characterizes the intuitive decision-making according to the CPT. This ontology is well-founded on the top-level ontology (UFO). Then, we developed a method using this ontology to automatically identify decision-makers' biased decisions due to risk seeking preference for losses, during decision-making. The ABI approach is implemented in the ABI tool, a computational tool that provides automatic support to detect and explain the risk seeking preference during the decision-making.

We focus on the identification of the risk seeking preference in decisions in the loss domain. More specifically, on GO/KILL decisions under risk that involve choosing between a sure loss and a higher probability of a larger loss, such as a portfolio manager deciding whether to kill a project that did not perform well or to invest more financial resources towards concluding it. However, it is beyond the scope of this research to eliminate cognitive biases and to analyze if the decision-maker decided rationally after receiving the risk seeking preference alert and its explanation from our ABI approach.

This research makes a novel contribution by automating the identification and explanation of risk seeking preferences using Cumulative Prospect theory from Behavioral Economics. The ABI approach transforms theoretical insights into actionable, real-time guidance, making them accessible to a broader range of organizations and decision-makers without requiring specialized personnel. By contextualizing CPT concepts into business language, the approach facilitates widespread adoption and enhances decision-making processes with deep behavioral insights.

Another significant contribution of the ABI approach is its ability to automatically collect historical data on risk seeking preferences and decision outcomes. This dataset provides valuable insights for strategic management and offers researchers opportunities to study the effects of bias awareness on decision-making (SAYÃO and BAIÃO, 2023).

To evaluate the ABI approach, we did an experiment with 193 participants. This experiment provided evidence indicating that the use of the ABI approach may contribute to organizational decision-makers to identify that they are subjected to risk seeking preference during the decision-making process, thus being a promising solution to support decision-makers within organizations and elsewhere.

The remainder of this paper is structured as follows. Section 2 presents the background of this research. Section 3 introduces the methodology and related works. Section 4 shows the ABI approach, and section 5 points to its evaluation. Finally, Section 6 concludes this work.

## 2 BACKGROUND

Decisions can be made rationally, studied by rational decision-making theories, such as Expected Value theory (EVT) and Expected Utility theory (EUT), or intuitively, studied by Behavioral Economics (BE), such as the Cumulative Prospect theory (CPT). According to rational decision-making theories, in situations of risk and uncertainty, the decision-maker should prefer the option with the greatest expected desirability or value/utility (STEELE and STEFÁNSSON, 2015).

Section 2.1 presents the definition of decision considered in this research, section 2.2, the decision-making-process, and the basic concepts related to Rational and Intuitive Decision-Making are presented in sections 2.3 and 2.4 respectively, mainly about EVT and CPT that are important concepts considered in the ABI approach.

### 2.1 Decision

Organizations - whether private, public, business enterprises, large government agencies, labor unions, large hospitals or large universities - face complex and uncertain environments and are under pressures that force them to respond quickly to changing conditions and to be innovative (DRUCKER, 2006)(KHANDELWAL and TANEJA, 2010)(SHARDA et al., 2014).

The most crucial element of organizations is people and their effectiveness in handling problems depends "as heavily on the effectiveness of the thinking, problem-solving, and decision-making that people do as upon the operation of the computers and their programs" (SIMON, 1997, p. 227). "All the decision-maker`s behavior involves conscious or unconscious selection of particular actions out of all those which are physically possible to the actor and to those persons over whom he exercises influence and authority" (SIMON, 1997, p.3).

Decisions may be made under risk and uncertainty. Knight (1921, p.233) defines risk as a measurable probability while uncertainty is defined as a probability that cannot be measured. In the case of risk, "the distribution of the outcome in a group of instances is known (either through calculation a priori or from statistics of past experience), while in the case of uncertainty this is not true, the reason being in general that it is impossible to form a group of instances, because the situation dealt with is in a high degree unique" (KNIGHT, 1921, p.233).

In this research, we specifically focus on decision-making under risk, addressing risks that are quantifiable, or estimable or probable. In decision-making under risk, decision-makers evaluate multiple potential outcomes, each with its probability. They use these probabilities to assess the risks and make choices, often aiming to maximize expected outcomes based on these probabilities, such as using the Expected Value theory, which is presented in section 2.3. This approach is commonly used in scenarios where past data or statistical models provide a basis for calculating the likelihood of each outcome.

Moreover, this research focuses on GO/KILL decisions, which involve deciding to keep or to cancel an ongoing project. As an example, deciding to keep or remove an ongoing project from the portfolio, which is a collection of projects or programs that are being undertaken by the organization to achieve its strategic objectives. This kind of decision is far from trivial and is characterized by uncertain and changing information, dynamic opportunities, multiple goals and strategic considerations and interdependence among projects (LEVINE, 2005). Moreover, with this kind of decisions' situations, decision-makers may have a bias toward risk seeking preference or risk averse preference.

More specifically, we focus on individual GO/KILL decisions under risk, associated with an organization goal, with no time pressure and involving high losses of medium and high probability, because in this context, the decision-maker often relies on intuition and may be influenced by cognitive biases (KAHNEMAN et al., 2021)(SADLER-SMITH, 2019). These cognitive biases may lead the decision-maker to have risk seeking preferences and make decisions that can cause high financial negative impact on the organizations (KAHNEMAN, 2011).

In the next section we discuss the decision-making process, as understood in the context of this proposal.

### 2.2 Decision-Making Process

Most strategic organization decisions take time, and it is therefore natural to divide them into phases (HANSSON, 2005). Organizational decision-makers usually make decisions by following the four-phase process of Simon (SHARDA et al., 2014) as illustrated in figure 1: intelligence, design, choice, and implementation.

The cycle of phases is far more complex than this sequence suggests. Each phase is itself a complex decision-making process. Moreover, activities of different phases may occur in parallel and at any phase may be a return to a previous phase (SIMON, 1977). Below, we show the activities of each phase, according to Simon (1977 apud SHARDA et al., 2014).

In the intelligence phase:
(1) define the organization objectives
(2) identify the problem

(3) define the problem
(4) identify the problem ownership
In the design phase:
(5) formulate a model: simplify reality and write down the relationships among all the variables
(6) identify criteria for choice
(7) identify alternative courses of action (possible solutions to the problem)
(8) predict and measure outcomes
In the choice phase:
(9) assess each alternative
(10) compare the alternatives of solution
(11) choose an alternative
(12) plan for implementation
In the implementation phase:
(13) put the recommended solution to work
(14) collect and analyze data to learn from the previous decisions and improve the next decision
(15) evaluate the implementation of the solution. Failure leads to a return to an earlier phase of the process. Successful implementation results in solving the real problem.

In this research, our focus is on the choice phase, more specifically, in activities 9, 10 and 11, where the alternatives are evaluated, compared and a final choice is made. In activity 11, the goal is to choose the alternative with the highest perceived value (BAZERMAN and MOORE, 2013). The decision is made here. The choice may represent the best or a good-enough solution, but also may be a biased decision.

In this research, we consider that all previous activities to activities 9, 10 and 11 were correctly executed without bias, i.e., the alternatives were identified as well as the financial impact and the risk probability of each one of them.

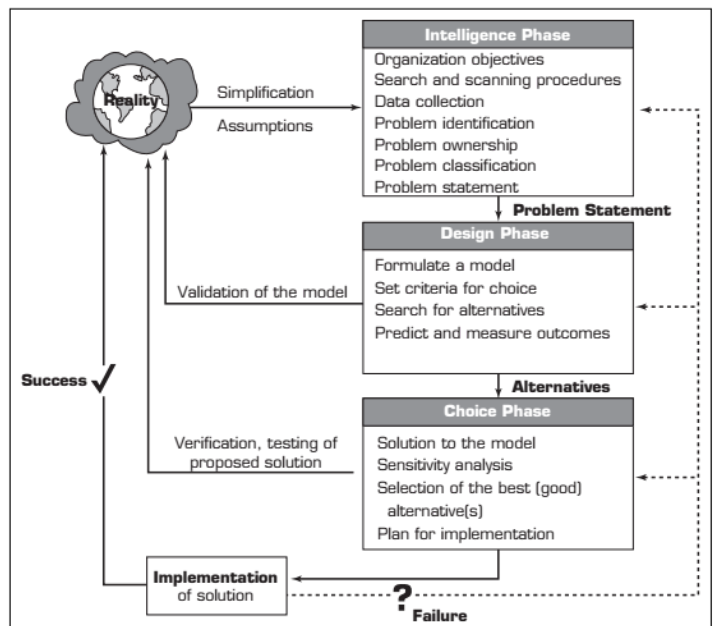

Figure 1: Simon's four-phase model of Decision-Making (1977 apud SHARDA et al., 2014, p.73).

Decisions can be made rationally or intuitively and are explained in sections 2.3 and 2.4, respectively.

## 2.3 Rational Decision-Making

The rational decision-making theory assumes that organizational decision-makers seek to use information rationally to make decisions to pursue their organization goals (FRENCH et al., 2009).

The term rationality refers to the "decision-making process that is logically expected to lead to the optimal result, given an accurate assessment of the decision-maker's values and risk preferences" (BAZERMAN and MOORE, 2013, p. 5). According to rational decision-making theory, such as Expected Utility theory (EUT) and Expected Value theory (EVT), in situations of uncertainty, the decision-maker should prefer the option with the greatest expected desirability or value (STEELE and STEFÁNSSON, 2015). In this case, the chosen alternative is demonstrably the best of all possible alternatives: the optimized alternative (SHARDA et al., 2014). Decision makers optimize expected values based on rigorous normative conditions rooted in mathematical logic (ARNOTT and GAO, 2022).

For simplicity of application, in this research we consider the EV theory. Expected Value theory can be traced to a correspondence in 1654 by Fermat and Pascal that laid the mathematical foundation for probability theory (FOX et al., 2015). The "expected" term is used because the mathematical concept is called expectation. Expected Value is a weighted average of the possible outcomes, and each outcome is weighted by its probability. So, the EV of 80% chance to win $100 and 20% chance to win $10 is $82. It is calculated by the sum of the products of the probability and outcome: (0.8 × 100 + 0.2 × 10) (KAHNEMAN, 2011). The EV can be calculated by the formula (1):

$$EV(x) = \sum_{i=1}^{n} p_i x_i \qquad (1)$$

where xi represents values of the possible outcomes of the random variable X, such as x1, x2, …, and pi represents their corresponding probabilities, such as p1, p2, etc.

Expected Value considers that people are rational and always choose options that have the highest Expected Value (Expected Value Maximization). For example, consider the following alternatives a decision-maker is facing:

According to the EVT, in decision-making under risk, decision-makers should always choose option B because its EV is higher than the EV of option A. However, in some contexts, EV is inadequate in explaining observed behavior because it does not consider risk preferences of decision-makers and cannot explain why people buy lottery insurance or tickets, or why, in the example above, most decision-makers would choose option A, the sure gain, instead of option B, that has the highest EV.

In the next section, we present Behavioral Economics and its main theory, the CPT, that was proposed as a response to the major violations of rational decision-making theories in decisions under risks.

## 2.4 Behavioral Economics and Intuitive Decision-Making

Behavioral Economics (BE) is a descriptive approach within Behavioral Decision theory that seeks to understand the factors influencing how and why individuals make decisions. BE is not a single reference theory, it is a complex web of interacting theories, phenomena, methods, and processes. Unlike traditional neo-classical economics, which assumes perfect rationality in decision-making, BE acknowledges that humans often deviate from rationality. It explores the cognitive biases, heuristics, and social influences that impact decision-making in economic contexts. By relaxing the assumption of perfect rationality, BE provides insights into the complexities of human behavior and offers a more realistic understanding of decision-making processes (ARNOTT and GAO, 2022; THALER, 2015).

In recent decades, BE has become increasingly significant in the field of decision-making research, largely due to the contributions of Nobel laureates Herbert A. Simon, Daniel Kahneman, and Richard Thaler. Many of the concepts related to decision-making in this research are based mainly on the work of Daniel Kahneman and his collaborator Amos Tversky.

It is important that decision-makers understand their own decision-making processes in order to clarify where they are likely to make mistakes influenced by cognitive biases. However, decision-makers "do not understand how they actually make decisions" (BAZERMAN and MOORE, 2013, p.5). In this sections, we present the concepts of intuitive decision-making and its cognitive biases, the Cumulative Prospect theory, and challenges and conclusion.

*2.4.1 Intuition.*

Research in managerial decision-making has shown that experienced managers resort extensively to intuition to make decisions, mainly under conditions of dynamism, uncertainty, and time pressure (KAHNEMAN et al., 2021)(SADLER-SMITH, 2019). Intuition is supposed to help decision-makers address complex problems (RIEL and HORVÁTH, 2014) and is a viable basis for experienced managers to make organizational decisions under these conditions (KAHNEMAN et al., 2021)(SADLER-SMITH, 2019).

One of the main foundations of BE is the Dual Process theory of human cognition. It posits that human thinking and decision-making processes are governed by two distinct systems: intuitive system (system 1) and rational system (system 2). The intuitive system represents intuitive and automatic thinking, characterized by quick and effortless processing of information. It operates on heuristics, biases, and associations, relying on past experiences and patterns to make rapid judgments. In contrast, the rational system reflects deliberate and conscious thinking, involving slow and effortful processing. It engages in logical reasoning, calculations, and analytical thought processes. The Dual Process theory posits that decision-making takes place within and between these two cognitive systems (KAHNEMAN, 2011; KAHNEMAN and KLEIN, 2009) (JARRAHI, 2018)(SIMON, 1987).

On one hand, the intuitive system is the origin of most of what we do right but on the other hand, the intuitive system is also the origin of much that we do wrong. The intuitive answers originate from skills or from heuristics, and they come to mind quickly and confidently. The intuitive system rarely checks its created answers, and this is the source of predictable biases (KAHNEMAN 2011).

The heuristics and bias stream of BE research is mainly devoted to understanding intuitive system decision processes, and its successes and failures (ARNOTT and GAO 2019). The next section presents the definition of cognitive biases.

*2.4.2 Cognitive Bias.*

Organizations can be negatively impacted by errors in judgment and decision-making. Bias and noise are distinct components of error. Bias is a systematic error we can often see and even explain; thus, we can predict. On the other hand, noise is random and, thus, it is an unpredictable error that we cannot easily see or explain (KAHNEMAN et al. 2016; 2021). Even experienced organizational decision-makers are also subject to biases when they think intuitively, since heuristics and skill are alternative sources of intuitive judgments and choices (KAHNEMAN, 2011). Our focus in this research is on bias.

Cognitive bias is also called psychological bias (KAHNEMAN et al. 2021). In this sense, a cognitive bias may be defined as "a systematic (non-random) error in thinking, in the sense that a judgment deviates from what would be considered desirable from the perspective of accepted norms or correct in terms of formal logic" (ARIELY, 2008). A cognitive bias is a systematic deviation from Expected Value. It is a predictable error of judgment that recurs predictably in particular circumstances (KAHNEMAN, 2011). Despite its recurrence, most decision-makers are rarely aware of their own biases when they are being misled by them. It is difficult for them to recognize their own biases (KAHNEMAN et al., 2021; LUCA and BAZERMAN, 2021).

When making decisions under risk and uncertainty, where one of the alternatives involves a sure high financial loss and the other option involves a higher financial loss of medium to large probabilities, decision-makers tend to reject the favorable alternative (sure high loss) and choose the unfavorable one. In this context, their behavior is influenced by a cognitive bias (loss aversion), that sets their preferences to risk seeking (KAHNEMAN and TVERSKY, 1979; TVERSKY and KAHNEMAN, 1992), hoping to avoid the sure loss. Decisions influenced by the risk seeking preference can lead to severe negative consequences to the organization. Hence, in this research, we focus on the risk seeking preference for losses of medium and high probability.

Loss aversion is a well-known cognitive bias in which a loss hurts more than an equivalent gain gives pleasure (KAHNEMAN, 2011; THALER, 2015). We tend to avoid losses more strongly than to have gains (KAHNEMAN, 2011). When decision-makers face very bad options, i.e., when offered a choice between a sure large loss and a gamble for a larger loss of medium and high probability, they hope to avoid the sure loss and become risk seeking. Loss aversion is utilized to explain other biases, such as the sunk cost fallacy (SAMSON, 2023).

The sunk cost fallacy, also known as escalation of commitment, refers to the tendency to continue investing in a decision based on cumulative past investments (time, money or effort), despite new evidence suggesting it may be wrong (ARKES and BLUMER, 1985; FLYVBJERG, 2021). This bias represents a non-rational commitment to an action (ARNOT and GAO, 2022) and involves factoring in non-recoverable historical costs when considering future actions (LOVALLO and SIBONY, 2010). It occurs when decision-makers make current choices based on past investments that no longer have relevance to the present situation (KAHNEMAN, 2011). The sunk cost fallacy is recognized as one of the ten most significant biases impacting project planning and management (FLYVBJERG, 2021).

In this research, we focus on risk seeking behavior for losses. Our proposed ABI approach identifies this context of decision-makers' risk seeking preference independently of the cognitive biases that caused the risk behavior, and explains the risk seeking preference according to the CPT, whose loss aversion bias is a key concept. In the experiments to evaluate our proposed ABI approach, we used decision-problems based on the sunk cost bias because it triggers the risk seeking preference when dealing with losses.

Debiasing is "a process whereby the negative consequences of a cognitive bias are reduced or mitigated" (ARNOTT and GAO 2022, p. 88). Cognitive debiasing involves a succession of stages, it starts with not knowing about biases, then realizing they exist, learning to identify them, thinking about making changes, deciding to make those changes, starting to use strategies to accomplish change, and finally, maintaining those changes (CROSKERRY et al., 2013). Identifying the cognitive bias is an important step of many approaches to removing or reducing the negative bias impact, i.e. debiasing (TSIPURSKY, 2019; BAZERMAN and MOORE, 2013; KAHNEMAN, 2011). For successful debiasing, it is crucial that decision-makers are committed to fighting the biases (KAHNEMAN et al., 2021).

In the IS domain, the debiasing strategy of Arnott (2006) is constituted by the following steps:
1- Identify the existence and nature of the potential bias.

2- Identify the likely impact and the magnitude of the bias.
3- Consider alternative means for reducing or eliminating the bias.
4- Reassure the user that the presence of biases is not a criticism of their cognitive abilities.
The focus of this research is on the first step of the above debiasing process.

In order to improve the decision-makers' decisions, Kahneman et al. (2016; 2021) propose the following corrective actions to cognitive bias:

- training decision-makers to detect situations in which cognitive biases are likely to occur.
- criticizing important decision, focusing on likely biases.
- asking a designated decision observer (outside human person) to search for diagnostic signs that could indicate, in real time, that other decision-makers are being affected by specific biases, such as sunk cost. The observer uses a checklist to search for these diagnostic signs. This approach is called "bias observation checklist" (KAHNEMAN et al., 2021). This approach is particularly applicable to group decisions but are applicable to individual decisions as well.

Educating people to overcome their biases is hard and more challenging than it seems. Moreover, it is difficult to know exactly which cognitive biases are affecting a judgment (KAHNEMAN et al. 2021). Criticizing the decision focusing on likely biases may help identify biases, but there is the problem that it is difficult to be aware of our own biases. Therefore, asking a designated decision observer to identify specific biases in the decision-makers' decisions during their decisions is the most promising technique according to Kahneman et al. (2021).To be effective in the "bias observation checklist" approach, decision observers need some training and tools and one of these tools is a checklist designed to identify key biases. Checklists are proven to enhance decision-making, especially in critical situations, by avoiding repeat errors. A checklist is not comprehensive of all possible biases; rather, it focuses on the most common and significant ones relevant to an organization's decisions. The bias observation checklist must be developed and customized according to the needs of the organization. Using such checklists when observing decisions can help reduce bias impact (KAHNEMAN et al., 2021).

As the focus of this research is on risk seeking preference for losses, we are considering that the "bias observation checklist" approach uses the following question which checks for sunk cost fallacy: "Are the people making the recommendation overly attached to past decisions?" (KAHNEMAN et al. 2011, p.57). This question is the question 9 of a tool created by Kahneman et al. (2011b), based on a 12-question checklist, that is intended to identify cognitive biases of the team making recommendations to the final decision-maker.

The "bias observation checklist" is a good approach because it is difficult for decision-makers to find biases in themselves and it is easier to identify bias in others (KAHNEMAN et al., 2021; 2011; 2011b). However, some drawbacks are that it depends on another person who should be trained to be the designated decision observer and that should not be susceptible to the cognitive biases of their own (KAHNEMAN et al., 2021). So, the bias check list process depends on the subjectivity of the decision observer, and as a human being, they may be affected by cognitive biases as well. Therefore, this person may negatively impact the process of bias identification. Another drawback is that "a decision observer is not an easy role to play, and no doubt, in some organizations it is not realistic" (KAHNEMAN et al., 2021, p.240).

Our solution proposal (ABI-Automatic Bias Identification) shares the key principle of the "bias observation checklist" approach that is the outside observer, but instead of being a human observer we propose a computational tool to automate the identification of the risk seeking preference during the decision-making. The ABI approach, while currently focused on identifying risk seeking preference and the loss aversion bias in loss domains involving losses of medium and high probabilities, has been designed with the flexibility to encompass a broader range of cognitive biases and risk preferences.

The ABI approach is applied to organizational decision-making and is independent of the human observer. It acts like an automated observer but has the advantage of not depending on another person to be the designated decision observer and the advantage of not being affected by human observer's cognitive biases.

In the next section, we present Cumulative Prospect theory, which is the most famous theory of Behavioral Economics. It is based on important concepts, such as loss aversion, and is the foundation of the ABI approach.

*2.4.3 Cumulative Prospect Theory.*

CPT is a descriptive theory of decision-making in risky and uncertain situations (TVERSKY and KAHNEMAN, 1992). It considers the fact that, in general, people tend to measure uncertainty and risk badly (SHARDA et al., 2014). CPT is an alternative to the rational decision-making theories and has no pretension to be a Normative theory guiding the rational choice. Instead, CPT is a theory about decision-makers' behavior and gives a good prediction of the actual choices real people make (THALER, 2015).

Cumulative Prospect theory extended Prospect theory (KAHNEMAN and TVERSKY, 1979) to uncertain as well as to risky prospects with any number of outcomes while preserving most of its essential features. Moreover, CPT allows different weighting functions for gains and for losses. Prospect theory was developed from empirical experiments on human choice behavior to provide an accurate descriptive model of decision-making. The choice problems of the experiments that laid the groundwork for Prospect theory were designed with alternatives that were structured to be clear and unambiguous, facilitating easier evaluation. Kahneman and Tversky (1979; 1992) utilized decision problems in their experiments consisting of two alternatives presenting just financial values and their associated risks, in order to gain more control over the experimental environment. This approach was key to their experiments, ensuring a controlled setup that allowed the specific observation of risk preferences and biases in decision-making.

CPT states that decision-makers under risk and uncertainty often do not make decisions in a rational way, such as calculating the Expected Value, as their decisions deviate from maximization. To understand it, we here compare it to the Expected Value theory. EVT assumes that decision-makers optimize by choosing the alternative with the highest Expected Value among all the alternatives.

The EVT and CPT predict differently the decision-maker's choice between Gamble A and Gamble B in the following example. Gamble A represents a sure loss of $1000, and Gamble B, 50% chance of losing $2500 or 50% chance of losing nothing. The EV of Gamble A is -$1000, and the EV for Gamble B is -$1250. According to EV, Gamble A would be selected as the final decision because they lose less (-$1000) than selecting the other option (-$1250). However, CPT argues the decision-maker chooses Gamble B. The key to understanding this example is that CPT considers the notion of reference point and loss aversion in decision-making, while EV does not. We show this example in detail in Section 4.2, but first we explain the main characteristics of Cumulative Prospect theory.

The following three cognitive features, all operating characteristics of System 1 (Intuitive), are the foundation of CPT and are illustrated as the CPT value function in Figure 2: (i) Reference dependence: evaluation is relative to a neutral reference point. Outcomes that are better than the reference point are seen as gains and are seen as losses when they are below the reference point. A reference point may be the status quo, but it can also be the outcome to which you feel entitled to, or the outcome you expect, or a goal in the future, for instance. In the aforementioned example, the reference point was "losing nothing" in the Gamble B. So, not achieving the goal is a loss; (ii) Loss aversion: refers to the tendency of individuals to experience greater psychological impact from losses than equivalent gains, i.e. a loss hurts more than an equivalent gain gives pleasure (THALER, 2015). We tend to avoid losses more strongly than to have gains (KAHNEMAN, 2011), that is why Gamble A is perceived as a loss. The thought of accepting the sure large loss is too painful. It has become the most powerful tool in the behavioral economist's arsenal (THALER, 2015); (iii) We feel diminishing sensitivity to gains and losses (THALER, 2015). We perceive the subjective difference between $1100 and $1200 as much smaller than the difference between $100 and $200 (KAHNEMAN, 2011).

In addition, probability weighting is another important component of Cumulative Prospect theory: the weighting function overweights low probabilities and underweights high probabilities. For instance, in a rational choice, the decision weight that corresponds to a 2% chance and 98% chance of winning a prize are 2 and 98 respectively. However, in CPT, the corresponding decision weights are 8.1 and 87.1 respectively (KAHNEMAN, 2011). Moreover, the qualitative improvement by increasing the chances from 0 to 5% is much higher than from 20 to 25% because in the first case it creates a possibility that did not exist earlier. This is called the possibility effect. Similarly, in the certainty effect the qualitative improvement from 95% to 100% has a large impact. In this case, outcomes that are almost certain are given less weight than their probability justifies (KAHNEMAN, 2011).

In CPT, decision-makers evaluate prospects in two phases: framing and valuation. In the framing phase, the decision-maker constructs a mental representation of the problem, considering relevant acts, contingencies, and outcomes. In the valuation phase, the decision-maker assesses the value of each prospect and makes choices based on these evaluations.

Figure 2 shows the psychological value, yield to a decision-maker, of gains and losses from a reference point, in the vertical axis. The monetary gains or losses of a decision are represented in the horizontal axis. The choice process of Cumulative Prospect theory has two phases: framing and valuation. In the framing phase, the decision-maker builds a mental representation of the problem, representing the acts, contingencies, and outcomes that are relevant to the decision. Then, in the valuation phase, they assess the value of each prospect and choose accordingly (TVERSKY and KAHNEMAN, 1992). One of the core achievements of CPT is the fourfold pattern of risk attitudes (TVERSKY and KAHNEMAN, 1992, p. 306).

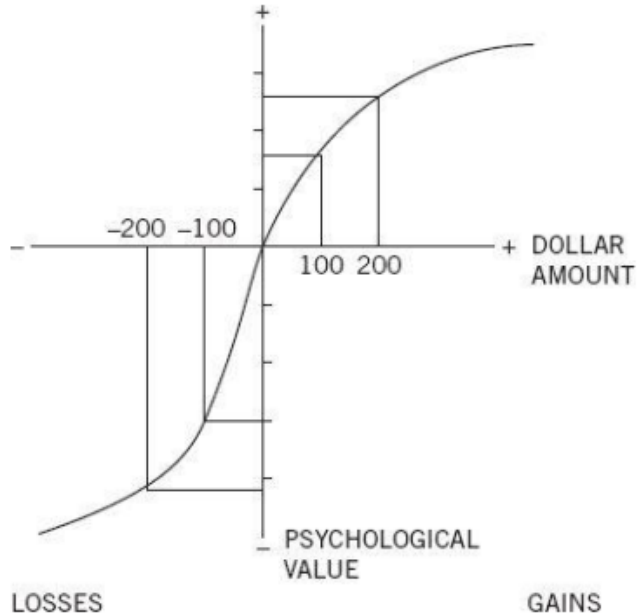

Figure 2: The form of the value function in CPT representing the psychological value of gains and losses (KAHNEMAN, 2011).

One of the key contributions of CPT is its fourfold pattern of risk attitudes, as shown in table 1. This pattern distinguishes risk aversion for gains and risk seeking for losses of medium and high probability, and risk seeking for gains and risk aversion for losses of low probability.

Table 1: The fourfold pattern of risk attitudes. Adapted from (KAHNEMAN, 2011).

| PROBABILITY | GAINS | LOSSES |
|---|---|---|
| HIGH PROBABILITY (Certainty effect) | "95% chance to win $10,000" or "$9,500 with certainty" Fear of disappointment. RISK AVERSE. Accept unfavorable settlement of 100% chance to obtain $9,500 | "95% chance to lose $10,000" or "$9,500 with certainty" Hope to avoid loss. RISK SEEKING. Reject favorable settlement and choose 95% chance to lose $10,000 |
| LOW PROBABILITY (Possibility effect) | "5% chance to win $10,000" or "$500 with certainty" Hope of large gain. RISK SEEKING. Reject favorable settlement and choose 5% chance to win $10,000 | "5% chance to lose $10,000" or "$500 with certainty" Fear of large loss. RISK AVERSE. Accept unfavorable settlement of 100% chance to lose $500 |

In this research, our focus is on the top right cell where the pattern of decision-makers' preference is risk seeking for losses of medium and high probability. When they face very bad options, i.e., when offered a choice between a sure large loss and a gamble for a larger loss of medium and high probability, they hope to avoid the sure loss and become risk seeking. They reject the favorable settlement of losing $9,500 with certainty and choose 95% chance to lose $10,000. The reasons for the risk seeking preference are: (i) there is diminishing sensitivity, i.e., the sure loss is very aversive and the pain of losing $9,500 with certainty is more than 95% of the pain of losing $10,000; (ii) the decision weight that corresponds to a probability of 95% to lose is only about 85, much lower than 95 (table 3), while the decision weight for a probability of 100% corresponds to 100. The risk seeking preference for losses was new and unexpected and is where many unfortunate human situations unfold. Decision makers tend to "accept a high probability of making things worse in exchange for a small hope of avoiding a large loss" (KAHNEMAN, 2011, p. 319).

For each of the cells of the fourfold pattern, systematic deviations from Expected Value are costly in the long run and this rule applies to both risk aversion and risk seeking (KAHNEMAN, 2011). Despite the influence of risk preferences on decision-making, many decision-makers are unaware of how these biases affect their choices, highlighting the need for increased awareness and understanding of the implications of CPT in decision-making processes.

*2.4.4 Challenges and Conclusion.*

Applying Prospect theory can be challenging because the specifics of its application are not always straightforward. This section discusses these challenges. Although this reseacrh concentrates on Cumulative Prospect theory, the challenges we outline here apply to both CPT and the original Prospect theory (KAHNEMAN and TVERSKY, 1979), since the former has been extended by the latter to include uncertain as well as risky prospects with any number of outcomes while preserving most of its essential features.

As the first challenge, some researchers argue that applying Prospect theory can be unclear because precisely defining gains and losses in specific situations is challenging, and identifying a suitable reference point can be problematic, which makes some scholars cautious about using the theory (BARBERIS, 2013). Despite this, various researchers have successfully adapted the concepts of gains and losses to their specific fields, such as Koszegi and Rabin (2006, 2007, 2009).

Second, Prospect theory is well-recognized for accurately describing risk attitudes in experimental environments. However, existing research has raised doubts about the reliability of its predictions in real-world situations beyond the laboratory (ABDELLAOUI et al. 2013; BARBERIS 2013; LIST 2003). Even so, there is evidence showing that PT predicts accurately in real-world situations (KACHELMEIER and SHEHATA 1992; POST, VAN DEN ASSEM, BALTUSSEN, and THALER 2008).

In addition, while real-world decisions surpass the complexity of binary financial choices with probabilistic outcomes, the application of PT within these controlled experiments offers a valuable framework for exploring how individuals confront risk and uncertainty, serving as a foundational guide to human decision behavior (BARBERIS, 2013)(ARNOTT and GAO, 2022).

Third, accurately estimating the probability of outcomes in actual organizational decision-making processes presents challenges for a number of reasons, such as the array of dimensions involved (including financial, technical, marketing, and production risks) (MARCH and SHAPIRA, 1987), the perceived probability of an event is subject to significant biases (TVERSKY and KAHNEMAN, 1974; KAHNEMAN and TVERSKY, 1979), and external factors further influence organizational decision-making (RAMOS et al., 2011), increasing the complexity of estimating probabilities. However, despite these challenges, research and practical methods support the estimation of probabilities (LOVALLO et al., 2020; MAIZLISH and HANDLER, 2005). Furthermore, key risk management guidelines—ISO 31000:2018 and COSO ERM: 2017—have recognized the importance of risk perception and the influence of cognitive biases in risk management (SIDORENKO, 2018).

Despite these challenges, the valuable insights and benefits that Prospect theory can provide in understanding organizational decision-making should not be underestimated. Thaler (2015) observes that Prospect theory is the seminal evidence-based theory in BE, and that PT provides a robust and comprehensive framework for predicting the actual choices that real people make in decision-making scenarios under risk. This is especially pertinent within organizations, where choices between alternatives often carry substantial implications.

Prospect theory is one of the most significant and widely cited works in Behavioral Economics, and indeed in the broader field of Economics. Although it is almost four decades old, PT remains compelling and insightful, as noted by Bernheim et al. (2018).

Ruggeri et al. (2020) successfully replicated Prospect theory in a multinational study with 4,099 participants from 19 countries and concluded that the principles of PT replicated beyond any reasonable thresholds. So, CPT also provides a useful framework for the descriptive analysis of decision-making under risk and uncertainty (TVERSKY and KAHNEMAN, 1992). Ruggeri et al. (2020) advocate for the continued investigation and application of PT's insights in diverse fields, ranging from financial decision-making to enhancing community health and well-being.

Even with the challenges, Prospect theory has been applied across a variety of economic contexts, influencing decisions related to consumer spending, labor engagement, and insurance policies (BARBERIS, 2013). Its impact extends beyond economics, infiltrating behavioral psychology and other fields such as Information Systems (ARNOTT and GAO, 2022) (ARNOTT and GAO, 2019) (SAMSON, 2023), health behaviors (GISBERT-PÉREZ et al., 2022), politics (THALER, 2015), business (EBRAHIMIGHAREHBAGHI et al., 2022; LERCHE and GELDERMANN, 2015), investment decision-making (BHATIA et al., 2021; KHILAR & SINGH, 2020; HERRMANN et al., 2015) and the banking industry (OGUNMOKUN et al., 2023).

The incorporation of Cumulative Prospect theory into this research, which involves the creation of the ABI approach and its implementation into an information system (the ABI tool) to assist decision-makers in identifying their risk seeking preferences, is supported by recent trends in Behavioral Economics within Information Systems research (ARNOTT and GAO, 2022). Given the emerging popularity of BE concepts like CPT and Debiasing methods, and their increasing relevance to IS research, this research aligns with the current trajectory of academic inquiry.

Behavioral Economics theories could be used in IS research to provide a better understanding of how organizational decision-makers make decisions and how to support them (ARNOTT and GAO, 2022). In this sense, the use of Cumulative Prospect theory is particularly pertinent considering the frequent occurrence of intuitive decision-making among organizational decision-makers, as identified by CPT. Since these quick, intuitive decisions are susceptible to biases, which lead them to have specific risk preferences, our suggested approach aims to counteract this by offering support grounded in CPT, thereby contributing to a more nuanced understanding of decision-making processes.

Moreover, Barberis (2013, p. 192) makes a reflection if evaluating risk according to CPT is a behavioral mistake and if there should be an effort to change people's behavior. As an approach to deal with this problem, he suggests "to explain to people, in an appropriate way, that they may be acting the way they are because of prospect theory preferences; and to then see if, armed with this information, they change their behavior". In this research, we are aligned with this approach in the sense that our ABI approach identifies the risk preference and explains it according to CPT.

Finally, the true semantics of what is value and how the cognitive biases and risk preferences may negatively affect decision-making are sometimes not well understood. The semantic notions involving these concepts, and their relations with decision-making can be better explored and represented, mainly by showing the differences between intuitive and rational decision.

## 2.5 Related Decision-Making Ontologies

The word "ontology" is used with different meanings in different communities. Two of the most prevalent uses in Computer Science literature refers to an ontology as "an explicit representation of a conceptualization (GRUBER, 1995)" and as "a formal, explicit specification of a shared conceptualization (STUDER et al., 1998)." Some of the main reasons to develop an ontology are: to share common understanding of the structure of information among people or software agents; to enable reuse of domain knowledge; to make domain assumptions explicit; to separate domain knowledge from the operational knowledge; and to analyze domain knowledge (NOY and MCGUINNESS, 2001).

A domain reference ontology (or simply, a domain ontology) "should be constructed with the sole objective of making the best possible description of the domain in reality w.r.t. to a certain level of granularity and viewpoint" (GUIZZARDI, 2005, p.37). In this research, we propose a domain ontology that is used as a basis for a computational decision-making tool.

Ideally, domain ontologies should be developed grounded in foundational ontologies (DE ALMEIDA FALBO, 2014). Our proposed ontology is a domain ontology grounded in UFO (Unified Foundational Ontology). In this section, we present the Unified Foundational Ontology (UFO), the Core Ontology on Decision-Making, and the Decision-Making Ontology (DMO), that were considered to propose our Intuitive Decision-Making Ontology. Moreover, we present some ontologies on cognitive biases.

UFO is a foundational ontology that has been developed based on various theories from areas such as Formal Ontology in philosophy, cognitive science, linguistics, and philosophical logics, and comprises a number of micro-theories about fundamental conceptual modeling notions.

The Unified Foundational Ontology (UFO) is divided into three layers dealing with different aspects of reality: UFO-A, UFO-B and UFO-C. UFO-A defines the core of UFO by discussing Endurants while UFO-B discusses Perdurants. Endurants are objects that persist over time preserving their identity, such as people and buildings. Perdurants, on the other hand, are composed of temporal parts over time and are understood as events, such as a process, a meeting, or a basketball match. Finally, UFO-C, that is based on UFO-A and UFO-B, is responsible for the discussion of social entities as Agents and their behavioral characteristics under an action. Figure 4 shows the categories around UFO-A and a few categories of UFO-B (GUIZZARDI et al., 2022).

Guizzardi et al. (2020) argue that for providing better support to decision-making, it is paramount to understand, first, the nature of decisions and of the decision-making process. For that, these authors proposed the Core Ontology on Decision-Making, founded on the Unified Foundational Ontology (UFO), on the ontology of Value Proposition (SALES et al. 2017) and on the ontological analysis of Economic Preference (PORELLO and GUIZZARDI 2018)(PORELLO et al. 2020). Our proposed ontology of this research reuses the concepts of the Core Ontology on Decision-Making and is described in Section 4.1.

The Decision-Making Ontology (DMO) proposed by Kornyshova and Deneckere (2010, 2011, 2012) models and formalizes decision-making in the Information System Engineering domain. This ontology has been developed aiming at: clarifying the concepts of the domain of decision-making, but also supporting the specification of decision-making requirements; and serving as a basis of a specification of the components of a DM method. It is part of an approach called MAke Decisions in Information Systems Engineering (MADISE).

The DMO was developed with the support of UML and defines explicitly the rational and the intuitive decision which are represented by the METHODBASEDDECISION and INTUITIVE DECISION classes respectively. Both classes specialize the DECISION class. Our proposed ontology reuses these 3 classes, but with different names: we call DELIBERATION in the place of METHODBASEDDECISION, INTUITIVE CHOICE in the place of INTUITIVE DECISION, and CHOICE in the place of DECISION. Neither this ontology nor the Core Ontology on Decision-Making consider that the decision-maker's preferences may be biased, thus influencing the Decision-Making process. We overcome this gap with our proposed Intuitive Decision-Making Ontology.

In relation to ontologies that may help understanding the semantic notions involving cognitive biases, risk seek preference, and their relations with decision-making, there is one ontology (LORTAL et al., 2014) used to reduce cognitive biases. The aim of this ontology is to quantify the risk of occurrence of a cognitive bias in the intelligence domain. In addition, Brodaric and Neuhaus (2020) provided a conceptual and formal foundation for an ontology of beliefs, desires, and intentions, and discussed how their theory can be extended to some major philosophical accounts of desires, and cognitive biases such as wishful thinking. However, the cognitive biases concerned by Tversky and Kahneman (1974) stem from the reliance on judgmental heuristics and are not attributable to motivational effects such as wishful thinking or the distortion of judgments by payoffs and penalties. None of these ontologies are related to risk seeking preference. Moreover, although there are other few proposals of ontologies (KORNYSHOVA and DENECKERE, 2010; NOWARA, 2011; NOWARA, 2017) that consider intuitive decision-making, which are the source of cognitive biases and risk preferences, they do not consider the cognitive biases neither the risk preferences.

In conclusion, to the best of our knowledge, there are no ontologies addressing relevant concepts to model the pattern of risk seeking attitudes as considered by the CPT for decision-making under risk and uncertainty.

## 3 METODOLOGY

This work follows the design science research (DSR) approach (VAISHNAVI and KUECHLER, 2004/2021), for solving a real problem using applied research in the field of Information Systems (HEVNER et al., 2004). In section 3.1, we present the systematic literature review we followed in this research. In section 3.2, we explain the main design science research (DSR) applied to our research.

### 3.1 Systematic Literature Review

This section presents a systematic literature review (SLR) of papers involving the topics of "organizational decision-making" and "risk seeking preference identification". In addition to the systematic review, we also have read several works that were referred to in this research (section 2), to have a better understanding on the topics related to decision-making under risk and uncertainty. To conduct the systematic literature review of papers, we used the guideline of Kitchenham and Charters (2007) that covers three phases of a systematic review: planning, conducting, and reporting the review.

*3.1.1 Review Question.*

As it is of interest to know how to identify the risk seeking preference during the organizational decision-making under risk, the review questions aim to investigate what the main approaches in this topic are. Thus, the questions below reflect the goals of this systematic review:

- What are the main approaches to identify the risk seeking preference during the organizational decision-making?
- What are the main advantages and disadvantages perceived in each type of approach to identify the risk seeking preference?

*3.1.2 Conducting Search.*

The search string was created by using the major terms of "identify", "risk seeking preference", "cognitive bias", "decision" and "organization". We used separate terms to have more results that could be of interest. We also have considered "loss aversion", "sunk cost", "escalation of commitment" and "diminishing sensitivity" as major terms because they are cognitive biases related to risk seeking preference. Moreover, we used the terms "reduce" and "debias" in order to see the papers about reducing cognitive biases because it can be related to the identification of the cognitive bias.

In addition, we have considered the alternate spelling and synonyms of the above terms. We have used them to form our search string by using the Boolean OR to concatenate synonyms and alternate spellings, and then by concatenating these terms by using Boolean AND to form the final search string as follows:

( TITLE-ABS-KEY ( ( "loss-aversion" OR "loss aversion" OR "risk preference" OR "risk propensity" OR "risk-seeking preference" OR "risk seeking preference" OR "seek risk preference" OR "risk-taking preference" OR "risk taking preference" OR "taking risk preference" OR "risk-seeking behavior" OR "risk seeking behavior" OR "seek risk behavior" OR "risk-taking behavior" OR "risk taking behavior" OR "taking risk behavior" OR "risk-seeking behaviour" OR "risk seeking behaviour" OR "seek risk behaviour" OR "risk-taking behaviour" OR "risk taking behaviour" OR "taking risk behaviour" OR "sunk cost" OR "sunk-cost" OR "scalation of commitment" OR "diminishing sensitivity" ) ) AND TITLE-ABS-KEY ( ( "decision" ) ) AND TITLE-ABS-KEY ( ( "reduce" OR "debias" OR "de-bias" OR "aware" OR "awareness" OR "identify" OR "identification" OR "identifying" ) ) AND TITLE-ABS-KEY ( ( "manager" OR "management" OR "finance" OR "organization" OR "investment" OR "organisation" OR " company" OR "enterprise" OR "business" OR "firm" OR "corporation" ) ) ).

We selected Scopus as the database for the search, because it delivers a broad coverage of any interdisciplinary abstract and citation database, greatly reducing the chances of missing key publications ( KITCHENHAM, 2010). Scopus indexes many of the leading publications, such as ACM, IEEE, and Elsevier publications. We focused on works published from 2015 to 2022. We chose 2015 as the starting year because the most recent foundation theory of Behavioral Economics (e.g., Prospect theory and Cognitive Biases) only recently has become predominantly used in decision support systems (DSS) research (ARNOTT and GAO, 2022), which is the field of Information Systems (IS) most concerned with decision-making. Arnott and Gao (2022) conducted the analysis in a sample period from 2014 to 2018. Surprisingly, it was lightly cited in DSS research until 2014 (ARNOTT and PERVAN, 2016; ARNOTT and PERVAN, 2014).

*3.1.3 Screening of Papers.*

In our systematic literature review, we applied specific inclusion (IC) and exclusion criteria (EC) to filter relevant studies addressing the identification of risk seeking preferences, loss aversion bias, sunk cost bias, or escalation of commitment bias, and also to filter relevant studies to enhance decision-makers' awareness of these biases or risk preferences.

Inclusion criteria focused on studies that defined, proposed, described, or evaluated approaches to identify risk seeking preferences or these biases, limited to journal or conference publications from 2015 to 2022. Exclusion criteria ruled out unrelated publications, incomplete papers, non-English works, and theses or dissertations.

The screening process, conducted from December 2021 and updated in May 2022, involved five steps: applying search strings, exclusion criteria, eliminating duplicates, reading abstracts, and full-text analysis. Out of an initial 347 studies, 8 relevant studies were selected for their alignment with our research objectives.

Among the relevant studies, we found that some approaches focused on the period before the decision-making process starts (EBRAHIMIGHAREHBAGHI et al., 2022; VOGEL & VOGEL, 2019; KHILAR and SINGH, 2020; SIROIS, 2019). For example, some approaches (KHILAR and SINGH, 2020; SIROIS, 2019) have an educational purpose, helping decision-makers identify that they are subjected to risk seeking preference by focusing on making them learn about it in advance. Additionally, some approaches helped decision-makers learn to select rational alternatives and avoid risky ones by observing previous rational decisions made by automated software agents (HERRMANN et al., 2015). However, none of these approaches supported decision-makers in identifying that they themselves were subjected to risk seeking preferences during the decision-making process.

We also found relevant studies with approaches that focus on during the decision-making process itself (BHATIA et al., 2021; OHLERT and WEIßENBERGER, 2020; LERCHE and GELDERMANN, 2015), all of which are automated by computational tools. Ohlert and Weißenberger (2020) focus on decision aids to reduce the sunk cost effect, and consequently, risk seeking preferences, finding that specific instructions for applying rational decision rules are more effective than simple warnings. Bhatia et al. (2021) investigate the impact of robot-advisory services on investment decision-making, aiming to mitigate cognitive biases such as overconfidence and loss aversion, recommending the best alternative and avoiding the one with risk seeking preference. Meanwhile, Lerche and Geldermann (2015) proposed the PT-PROMETHEE approach, utilizing Prospect theory and Multi-Criteria Decision Analysis to help decision-makers express loss aversion and enhance their preference awareness, though it has limitations in defining the reference alternative and calculating loss aversion coefficients. However, none of these studies offered a concrete mechanism for automatically identifying risk seeking preferences during decision-making.

Moreover, apart from Ohlert and Weißenberger (2020), most studies did not provide any explanation of these risk preferences to decision-makers, revealing a substantial gap in the literature on this subject. In addition, we highlight a lack of approaches using a formal representation of the knowledge related to the risk seeking preference, such as an ontology. This combined evidence highlights a critical need for an automated solution that not only identifies but also explains risk seeking preferences during decision-making (in real-time).

### 3.2 The Main Design Science Cycle of our Research

In this study, completed a main DSR cycle and two sub-DSR cycles, as illustrated in figure 3. Our research design consists of five steps: awareness of the problem, suggestion, development, evaluation, and conclusion.

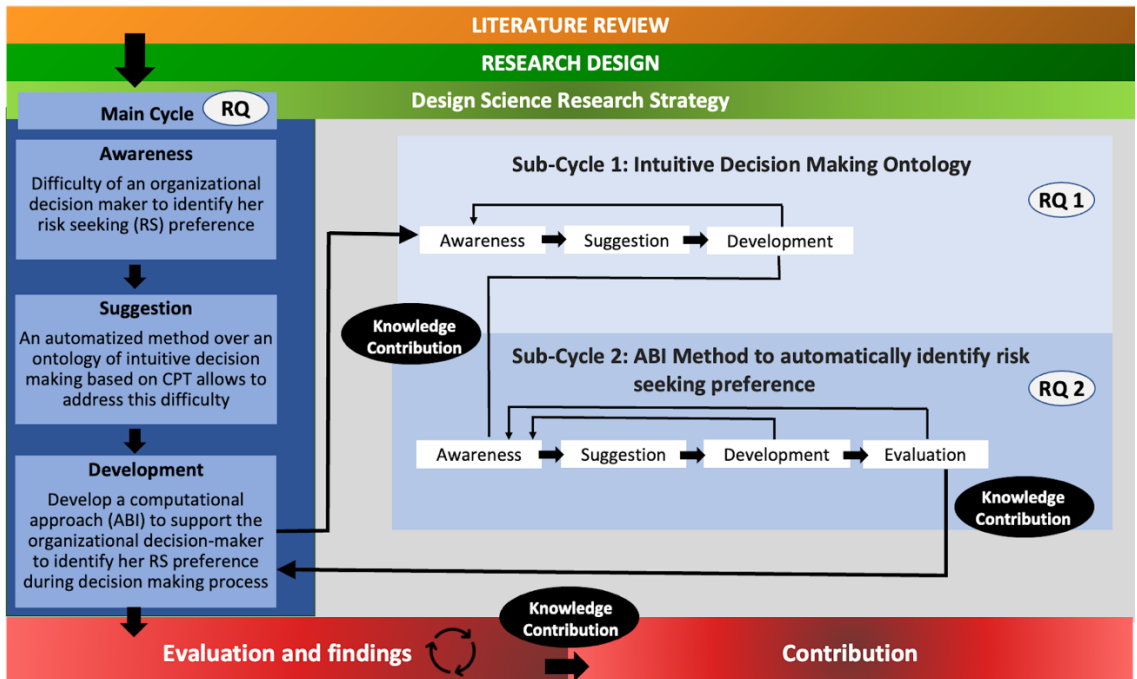

Figure 3: Design science research process followed in this research. Source: Prepared by the author (2024).

The awareness phase of the main DSR cycle was partly described in section 1 with the definition of the problem and one main research question (RQ). To answer the main RQ, we must answer the following and more specific RQ:

- RQ1: How to characterize an intuitive decision-making situation in which the decision-maker is subject to individual risk seeking preference?
- RQ2: How to identify that the decision-maker is subjected to individual risk seeking preference during the decision-making?

The results from this phase led to a suggestion to create an automated method over an ontology of intuitive decision-making based on CPT to address the research problem. This is the ABI approach which will identify and explain the risk seeking preference during decision-making. In this research, our proposed approach is the result of the analysis of the literature review presented on section 2 and on the SLR presented in this section: (i) intuitive decision-making and cognitive biases studied by Behavioral Economics - more specifically, by the Cumulative Prospect theory; (ii) decision-making ontologies; and (iii) approaches that support decision-makers to identify that they are subjected to risk seeking preference during decision-making.

The development phase of the main DSR cycle includes two sub-DSR cycles (figure 3) in order to develop an approach to identify and explain the risk seeking preference during decision-making. First, to be able to identify a situation that can lead to risk seeking preference we built an ontology to characterize the intuitive decision-making under risk and uncertainty in which decision-makers' decisions may be biased. This was done in sub-DSR cycle 1, which answers RQ1. This ontology can be used to identify but also to explain the reasons leading to risk seeking preference. The knowledge contribution of sub-DSR cycle1 led to sub-DSR cycle 2, which answers RQ2, to create a method to automatically identify risk seeking preference during decision-making and to explain to decision-makers their risk preferences and how their decisions may be biased. Sub-DSR cycle 2 also encompasses the conversion of this ontology to a relational database to be used by the method. In addition, the main development phase encompasses the implementation of ABI tool, that is the computational implementation of the ABI approach.

The evaluation of the ABI Approach, the artifact of the main DSR cycle, was done by an experiment. This is discussed in more detail in section 5.

The conclusion of the main DSR cycle occurred when the main research question was answered, and the knowledge contributions were documented. This is discussed in more detail in section 6. The main research question was answered with the development of an approach for automatic bias identification and explanation in decision-making under risk, named ABI Approach. To answer the main RQ, it was necessary to answer RQ1 and RQ2 first.

## 4 AUTOMATIC BIAS IDENTIFICATION (ABI) APPROACH

In this section, we present our solution proposal, called Automatic Bias Identification Approach (ABI Approach). In summary, the ABI approach supports the organizational decision-makers identifying their biased decisions during the choice phase of the decision-making process. More specifically, it focuses on the identification of the risk seeking preference.

The ABI (Automatic Bias Identification) approach consists of:
- an ontology to characterize intuitive decision-making under risk and uncertainty, which is explained in section 4.2.
- a method over this ontology to automatically identify decision-makers'risk seeking prference. The ABI method and the ABI process is explained in section 4.1. This method and all the ABI approach are implemented by the ABI tool that is explained in section 4.3.

### 4.1 ABI Process and ABI Method

Figure 4 shows an overall view of the ABI approach. In summary, it has 3 main steps:
1. Decision-maker inputs information about the decision problem and makes a decision.
2. The ABI approach automatically identifies if there is risk seek preference, and
3. It alerts the decision-makers with the bias explanation contextualized for the situation.

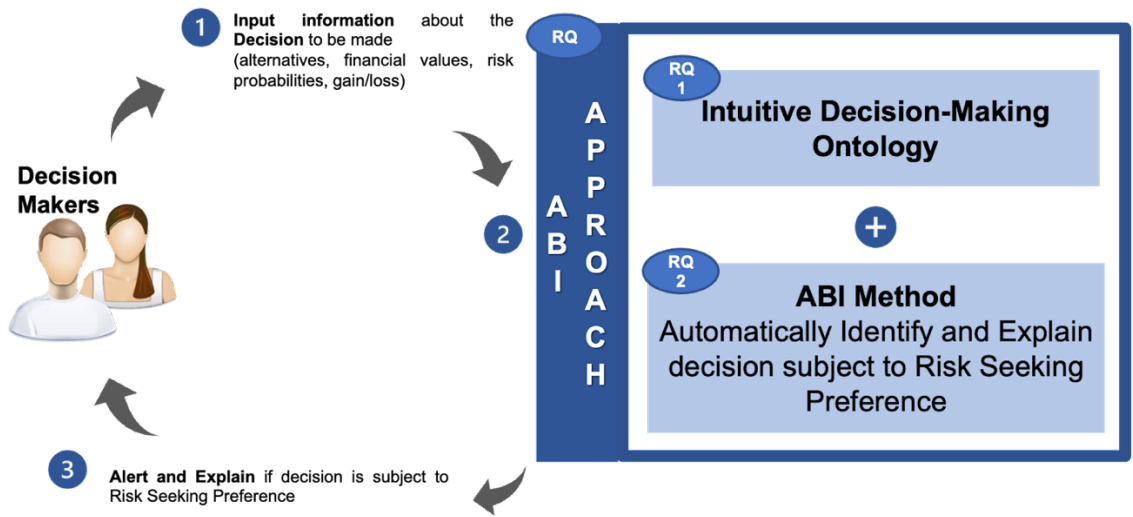

Figure 4: ABI Overview. Source: Prepared by the author (2024).

The ABI Approach is represented in more detail in the ABI Approach process of figure 5. First, the decision-maker informs the ABI approach which are the decision problem and the alternative choices. For each alternative, they inform the Financial Value and the Risk Probability. Next, they reason rationally or intuitively to make a choice in the sequence. This decision is made by choosing just one alternative.

Then, the ABI method receives as input the ontology to automatically identify if there is risk seeking preference. The activity "Identify Risk Seeking preference" performs a set of rules defined from the fourfold pattern of risk preferences of CPT and in the sequence, the ABI method instantiates the ontology. In this research, these rules were implemented in a Python function to identify the risk seeking preference for losses of medium and high probability, which are explained in section 4.3.2. This activity returns TRUE and the explanation of the risk seeking preference contextualized for the situation when there is risk seeking for losses in the selected choice or FALSE when there is not. These outputs go to the decision-maker as an alert but can also go to a Debias process that will use it as a basis for a strategy to debias the decision-maker.

Figure 5 illustrates a general Debias process that receives the output of the ABI approach and uses it in a strategy to reduce or eliminate the bias and, at the end of this process, gives the decision-maker the possibility to change their initial decision. In this case, the decision-maker reasons rationally or intuitively to make the choice in the sequence, by confirming the initial choice or by choosing another alternative. The process in figures 5 was modeled by Bizagi Process Modeler software (BIZAGI, 2022) using BPMN 2.0 notation (OMG, 2022).

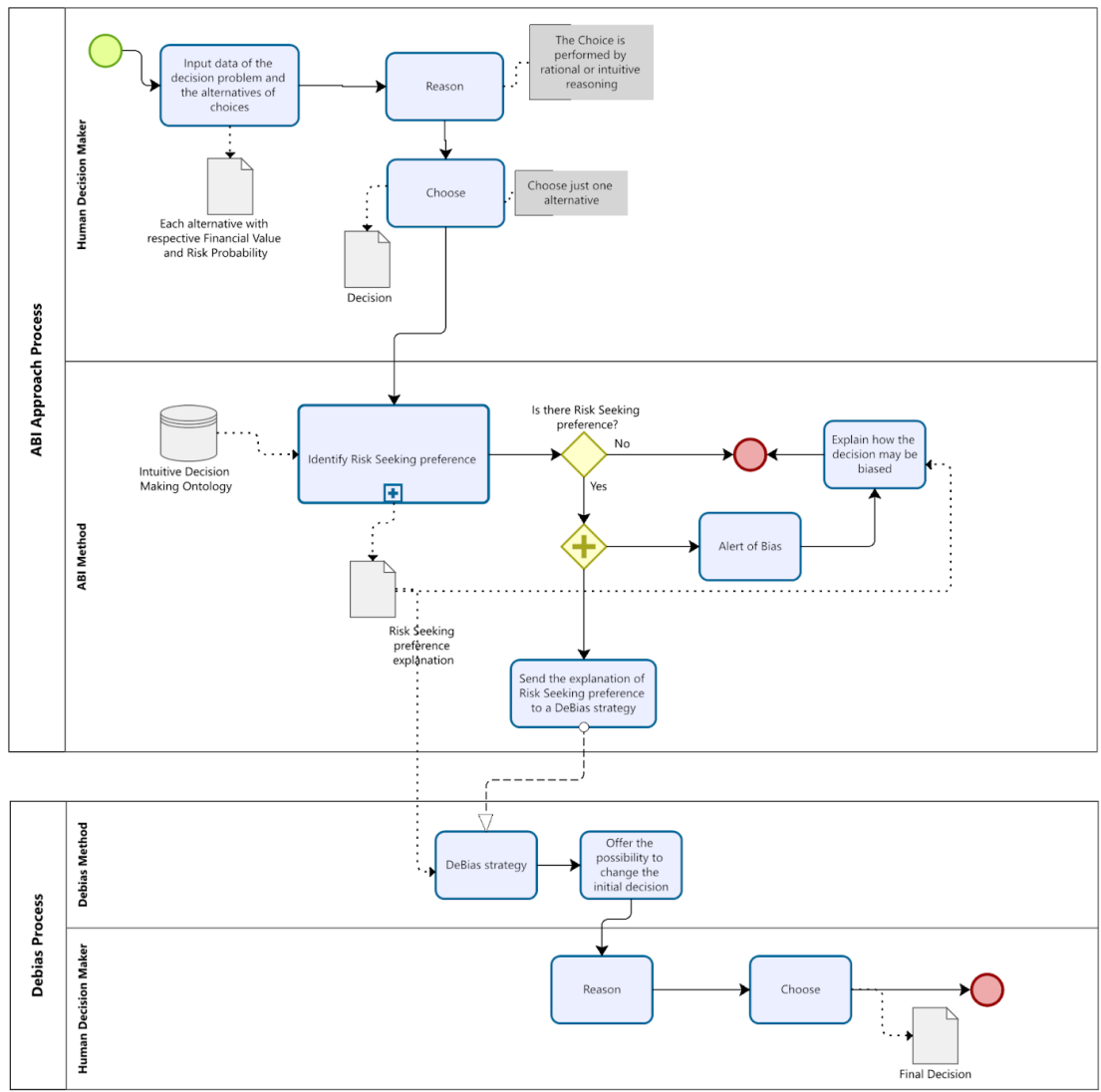

Figure 5: The ABI Process. Source: Prepared by the author (2024).

The ABI approach, while currently focused on identifying risk seeking preference in loss domains involving losses of medium and high probabilities, has been designed with the flexibility to encompass a broader range of cognitive biases and risk preferences. Future iterations could be adapted to recognize and address other biases and risk preferences, further enhancing its applicability and utility in decision-making processes.

### 4.2 Intuitive Decision-Making Ontology

The intuitive decision-making ontology (RAMOS et al., 2021) is the foundation of our ABI approach because it allows us to precisely characterize intuitive decision-making under risk and uncertainty and to explain how decisions may be biased (figure 6).

We built our ontology grounded on the Unified Foundational Ontology (UFO) (GUIZZARDI 2005, GUIZZARDI et al. 2015, GUIZZARDI et al. 2008). We extended the decision-making ontology of Guizzardi et al. (2020) by including intuitive decisions and associated concepts, according to CPT, and by using the classes DELIBERATION, INTUITIVE CHOICE, and CHOICE of the Decision-Making Ontology (KORNYSHOVA and DENECKERE, 2010; 2011; 2012). CPT considers decisions under risk and uncertainty, and so does our proposed ontology.

In Rational Decision theories, value (or use value) can be defined as a composition of benefits (which emerge from goal satisfaction) and sacrifices (which emerge from goal dissatisfaction) (LANNING and MICHAELS, 1998). This is the value concept which allows one to determine what are the alternatives (value bearers) to be chosen from, how they are valued, what are the applied criteria, who can execute the action resulting from the decision, and so on. In our proposed ontology, we call this value as RATIONAL VALUE to make the difference with PSYCHOLOGICAL VALUE explicit.

According to CPT, people attach values to gains and losses rather than to wealth (KAHNEMAN, 2011). Depending on how a decision-maker frames the decision problem (mental representation of the problem), she represents the value of the possible outcomes that may occur as gains and losses from a reference point. This is represented in our ontology as the PSYCHOLOGICAL VALUE.

Consider a SITUATION in which an AGENT must decide between two alternatives A and B. Each alternative is a VALUE BEARER (either a VALUE OBJECT or a VALUE EXPERIENCE). When an AGENT decides something, they can decide rationally or intuitively (KAHNEMAN, 2011) (KORNYSHOVA and DENECKERE, 2012). During the decision-making process, they take into consideration their own PREFERENCES regarding two possible VALUE BEARERS.

Let us now consider in more detail what happens when an agent makes a decision. The agent may decide rationally (i.e., performs a DELIBERATION) or intuitively (e.g., performs an INTUITIVE CHOICE). DELIBERATION (GUIZZARDI et al., 2020) and INTUITIVE CHOICE are manifestations of that agent's PREFERENCES over two VALUE BEARERS. Based on the Dual Process theory explained in Section 2, the VALUE ASCRIPTION can be PSYCHOLOGICAL or RATIONAL. If the AGENT deliberately assesses their options, then they are using Rational System , i.e., they are reasoning logically (a DELIBERATION is hence happening). This results in a RATIONAL VALUE ASCRIPTION (RVA). On the other hand, if the AGENT uses their intuition to decide, then they are using the Intuitive System, i.e., intuitive thinking. So, it is a PSYCHOLOGICAL VALUE ASCRIPTION (PVA). Note that the Rational and Intuitive Systems operate in parallel and interactively (KAHNEMAN, 2011; KAHNEMAN and KLEIN, 2009). To account for both systems, we incorporated the notion of a PVA in addition to a RVA, as described in the following paragraphs.

Figure 6: Intuitive decision-making ontology: deliberation, intuition, value, and preference (RAMOS et al., 2021).

Each RATIONAL VALUE ASCRIPTION is composed of several smaller "comparisons" (or "judgements"), named RATIONAL VALUE ASCRIPTION (RVA) COMPONENTS, which aggregate an INTENTION and INTRINSIC MOMENTS that are taken into consideration by the AGENT when ascribing VALUE to a VALUE BEARER. Each RVA COMPONENT is in its turn associated to a RATIONAL VALUE COMPONENT, defined as either a BENEFIT or a SACRIFICE.

Each PSYCHOLOGICAL VALUE ASCRIPTION is composed of several smaller "comparisons" (or "judgements"), named PSYCHOLOGICAL VALUE ASCRIPTION (PVA) COMPONENTS, which aggregate an INTENTION and INTRINSIC MOMENTS that are taken into consideration by the AGENT when ascribing VALUE to a VALUE BEARER. Each PVA COMPONENT is in its turn associated to a PSYCHOLOGICAL VALUE COMPONENT, defined as either a GAIN or a LOSS according to a REFERENCE POINT, which is ontologically a BELIEF. A reference point is highly determined by the objective status quo but is also affected by social and expectations comparisons. When an employee receives a smaller raise than everyone else in the office, they experience this objective improvement as a loss. Moreover, the PSYCHOLOGICAL VALUE COMPONENTS are influenced by the COGNITIVE BIASES, such as LOSS AVERSION, that is the DESIRE to avoid losses. For example, in many decisions, decision-makers must choose between retaining the status quo and accepting an alternative to it. Because losses loom larger than gains, they tend to be biased in favor of keeping the status quo, considering the status quo as the reference point (KAHNEMAN, 2011).

As seen before, the core premise of conventional economic theory is that people choose by optimizing (THALER, 2015). Hence, considering a RATIONAL VALUE ASCRIPTION, a VALUE BEARER is preferred in a has preference relation if and only if the value magnitude of its RATIONAL VALUE ASCRIPTION bearer is greater than the one of the compared alternatives (GUIZZARDI et al., 2020). However, the PSYCHOLOGICAL VALUE ASCRIPTION works differently. As shown by CPT, a VALUE BEARER that is preferred in a RVA may be the deprecated in a PVA because the PSYCHOLOGICAL VALUE is represented as gains and losses from a reference point rather than as final states of wealth, as assumed by EUT (FRENCH et al, 2009). This will set the PREFERENCE mode for most of the decision-makers as RISK SEEKING PREFERENCE or RISK AVERSE PREFERENCE. They tend to be risk averse for gains and risk seeking for losses of medium and high probability; risk seeking for gains and risk aversion for losses of low probability (KAHNEMAN, 2011).

From our ontology, we created a relational database schema to be used by the ABI tool.

**4.3 ABI Tool**

We implemented a tool to support the ABI Approach. In the development process, we initially did mockups. These mockups were used in the 6 pilot tests before the final experiment. In the sequence, we developed the ABI tool using Python (streamlit) as the programming language, and PostgreSQL 14.8 (POSTGRESQL, 2023) for managing the relational database. The ABI tool was created based on the intuitive decision-making ontology which we converted to a relational schema using PostgreSQL 14.8.

*4.3.1 ABI Tool Modules.*

The ABI tool is composed of 3 main modules. The Input data handler module, is where the human decision-maker inputs the main data about the decision, i.e., the decision problem and its alternative choices with the corresponding financial values and risk probabilities. In the Decision choice registration module, the human decision-maker decides by making a choice between the alternatives. Finally, in the Risk seeking preference identification module, the ABI tool identifies if the user has risk seeking preference and gives an alert, also including an explanation of the preference.

In a practical scenario, after the user, who can be the project officer, inputting the decision problem and its alternatives in the ABI tool (Module 1), the decision-maker makes a decision by making a choice between the alternatives (Module 2). If the ABI tool (Module 3) identifies risk seeking preference then it triggers an alert (figure 7 and figure 8) of the possibility of having risk seek preference and an explanation of the reasons. If there is risk seeking, the ABI method explains this behavior according to the ontology. It was implemented in the ABI tool by creating a customized image for the specific problem and decision at hand, as shown in figures 7 and 8. Figure 7 shows the first part of this alert which consists of explaining the purpose of the alert and recapping the decision problem and the choice made by the decision-maker.

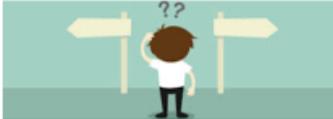

Figure 7: First part of the ABI tool alert (Module 3: Risk seeking preference identification). Source: Prepared by the author (2024).

Figure 8 shows the second part of this alert which consists of explaining the risk seeking preference according to the intuitive decision-making ontology and according to the context of the decision being made.

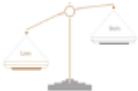

Figure 8: Second part of the ABI tool alert (Module 3). Source: Prepared by the author (2024).

*4.3.2 ABI Method Implementation to Identify Risk Seeking Preference.*

In this section, we present the rule used by the ABI method to automatically identify if there is risk seeking preference for losses of medium and high probabilities.

This rule was implemented in a python function called isRiskSeekingForLossesChoice(coproblem, codchoice) inside the ABI method. This function receives as parameters the problem code and code of the chosen alternative. In other words, this function knows the information about the alternatives, specifically financial values and risk probabilities, as well as the selected alternative as the final decision. This corresponds, respectively, to the VALUE BEARER and PREFERRED BEARER in the ontology (figure 6). Internally, this function calculates the Expected Value of each alternative, which corresponds to the RATIONAL VALUE ASCRIPTION in the ontology. Then, the function returns TRUE if there is risk seeking, or FALSE, otherwise.

In figure 9, we illustrated an example of decision problem used by the ABI method: 95% chance to lose $10,000 OR lose $9,500 with certainty. The data under the column "ABI method INPUT DATA" are the data used by isRiskSeekingForLossesChoice(), i.e., the features of alternatives 1 (in blue) and 2 (in green). The column "Did the ABI method identify option 2 is risk seeking for losses?" is the ABI method result.

First, we describe the features of alternatives 1 and 2, i.e., the input data for the ABI method. These descriptions are illustrated by using the instance of problem id 1 of figure 9, whose alternative 1 involves a sure loss and alternative 2, a high probability of a higher loss. Consider that the decision-maker chose option 2, which is the one related to risk seeking behavior, for id_problem 1. So, the isRiskSeekingForLossesChoice(id_problem, cod_option) function is called using 1 for the id_problem and 2 for the cod_option).

| | ABI method INPUT DATA | | | | | | | | | | ABI method RESULT |
|---|---|---|---|---|---|---|---|---|---|---|---|
| | OPTION 1 | | | | | OPTION 2 | | | | | |
| Problem id | Finantial value a1 | Probability a1 (%) | Finantialvalue b1 | Probability b1 (%) | EV1 | Finantialvalue a2 | Probability a2 (%) | Finantialvalue b2 | Probability b2 (%) | EV2 | Did the ABI method identify option 2 is risk seeking for losses? |
| 1 | -9500 | 100 | 0 | 0 | -9500 | -10000 | 95 | 0 | 5 | -9500 | Yes |

Figure 9: Example of decision problem used by the ABI method. Source: Prepared by the author (2024).

Features of alternative 1:

- Financial value a1: Financial value of a possible outcome of Alternative 1. For problem id 1, the financial value a1 is -9500.
- Probability a1 (%): Probability of the occurrence of Financial value a1. For problem id 1, the Probability a1 is 100%.
- Financial value b1: Financial value of the other possible outcome of Alternative 1. For problem id 1, the financial value b1 is 0.
- Probability b1 (%): Probability of the occurrence of Financial value b1. For problem id 1, the Probability b1 is 0.
- EV1: Expected Value of alternative 1. For problem id 1, EV1 is -9500.

Features of alternative 2:

- Financial value a2: Financial value of a possible outcome of Alternative 2. For problem id 1, the financial value a2 is -10000.
- Probability a2 (%): Probability of the occurrence of Financial value a2. For problem id 1, the Probability a2 is 95%.
- Financial value b2: Financial value of the other possible outcome of Alternative 2. For problem id 1, the financial value b2 is 0.
- Probability b2 (%): Probability of the occurrence of Financial value b2. For problem id 1, the Probability b2 is 5%.
- EV2: Expected Value of alternative 2. For problem id 1, EV2 is -9500, i.e., the same as EV1.

Here is the ABI method rule to identify if OPTION 2 has risk seeking preference for medium to high probabilities of losses:

***IF** ( (choice==OPTION2 and isProbability100(probabilitya1) and isNegativeValue(financialvaluea1) and isProbabilityHighAndLess100(probabilitya2) and isNegativeValue(financialvaluea2) and isNegativeOrZeroValue(financialvalueb2) and isAbsValuea2GreaterAbsValuea1(financialvaluea2, financialvaluea1) and isEV1GreaterEqualsEV2(EV1,EV2) ) **THEN** OPTION 2 has risk seeking preference for medium to high probabilities of losses.*

This rule is defined from the fourfold pattern of risk preferences of CPT, more specifically from risk seeking preference in the loss domain. If all predicates in the rule above are TRUE then there is risk seeking preference for medium to high probabilities of losses. Each predicate is explained below.

- choice==OPTION2: returns TRUE if the decision-maker chose the option 2. For problem id 1, it returns TRUE.
- isProbability100(probabilitya1): returns TRUE if probabilitya1 is 100%. For problem id 1, it returns TRUE.
- isNegativeValue(financialvaluea1): returns TRUE if financialvaluea1 is negative. For problem id 1, it returns TRUE.
- isProbabilityHighAndLess100(probabilitya2): returns TRUE if probabilitya2 is greater or equals than 50% and less than 100%. Kahneman and Tversky (1992) consider a high probability when it is >= 50%. For problem id 1, it returns TRUE.
- isNegativeValue(financialvaluea2): returns TRUE if financialvaluea2 is negative. For problem id 1, it returns TRUE.
- isNegativeOrZeroValue(financialvalueb2): returns TRUE if financialvalueb2 is negative or zero. For problem id 1, it returns TRUE.
- isAbsValuea2GreaterAbsValuea1(financialvaluea2, financialvaluea1): returns TRUE if the absolute value of financialvaluea2 is greater than the absolute value of financialvaluea1. For problem id 1, it returns TRUE.
- isEV1GreaterEqualsEV2(EV1,EV2): returns TRUE if EV1 is greater or equals EV2. For problem id 1, it returns TRUE, EV1 is equal to EV2.

In the next section, we present how the ABI tool was evaluated in the final experiment.

## 5 EVALUATION OF THE ABI APPROACH

The ABI approach was evaluated in a final experiment, addressing the main RQ and including some potential users of this tool. An experiment enables a researcher to test a hypothesized relationship between an independent variable (IV) and a dependent variable (DV) by manipulating the independent variable. The evaluation of the ABI tool included the evaluation of the intuitive decision-making ontology and the ABI method that consisted of having the computational ABI tool using the ontology and ABI method to identify and explain the risk seeking preference.

In this experiment, we used the within-subjects design. The within-subjects design, also known as repeated measures design, is a type of experimental design in which all participants in the sample are exposed to the same treatments, i.e., everyone in the sample takes part in every condition (STANGOR, 2014). There is just one group of participants that are tested more than once, and their scores are compared. The goal is to measure changes resulting from different treatments for outcomes or to measure changes over time. We used within-subjects because it minimizes the impact of individual differences, and it saves cost once it requires less participants than with between-subjects (STANGOR, 2014).

In order to keep the final experiment simple and at the same time relevant, we firstly did 6 pilot tests in the exploration phase. These pilot tests were necessary to test the experiment design in order to achieve its goals with the final and main experiment. In this section, we describe the final experiment which was done with the ABI tool by following the key steps involved in the experimental process presented below (MONTGOMERY, 2017).

### 5.1 Experiment Procedure and Material

In this study, participants were presented with a scenario where they were employees in a company and were tasked with recommending a choice between two alternatives that involved significant financial losses. The recommendation was intended for the decision-maker within the organization.

The experiment was conducted entirely online, ensuring no interaction between the participants. Written information was provided to all participants to avoid any potential bias that could arise from personal interaction with the researcher. This approach ensured that all participants received the same information and treatment. The experiment was conducted in Portuguese from Brazil, but here the figures have been translated into English.

From the researcher's perspective, the experiment consisted of three main steps, as illustrated in figure 10. From the participants' perspective, steps 1 and 2 were merged into a single step, resulting in a two-step process.

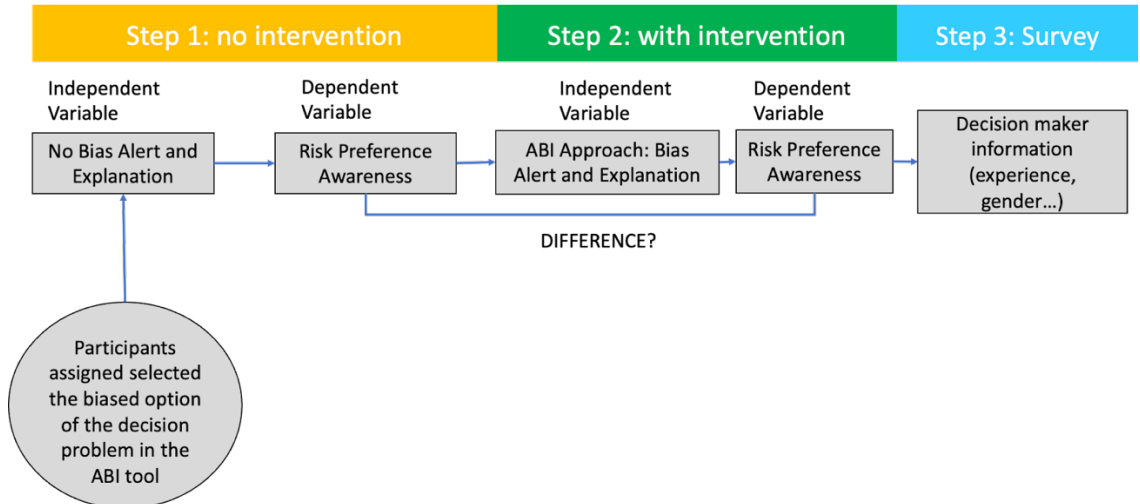

Figure 10: Final experiment design. Source: Prepared by the author (2023) and based on figure 10.4 of (STANGOR, 2014).

First, in step 1, participants provided informed consent and responded to a decision problem directly in the ABI tool (figure 11). Subsequently, their level of awareness regarding risk seeking preference was measured. This marked the conclusion of step 1, wherein there was no intervention involving the ABI approach, specifically no identification or explanation of the risk seeking preference.

**Decision Module**

Consider that you work in a small-sized company that has R$ 150,000 available for investment. The company has already invested R$ 200,000 in a project to create a new product in the market. On the initially planned date to complete the product development, the reported situation is that the project is behind schedule and would need an additional R$ 50,000 to complete the product. You need to make a recommendation to the decision-maker by choosing one of the two alternatives below. There are no other alternatives. Which alternative do you choose?

**Alternative 1**

1. Cancel the project and lose the R$ 200,000 already invested.

   In other words: 100% chance of losing R$ 200,000.

**Alternative 2**

2. Continue with the project by investing an additional R$ 50,000, even though there is a 90% chance of losing the entire investment (R$ 250,000) and a 10% chance of not losing anything.

   In other words: 90% chance of losing R$ 250,000 AND 10% chance of losing nothing.

Make a decision and enter the code of the chosen alternative (1 or 2) below. Then click the Save button:

0

Save

Figure 11: Final experiment: the ABI tool screen of the Decision Module with the decision problem. Source: Prepared by the author (2024).

The decision problem employed in this experiment (figure 11) followed a "typical structure in decision-making research: binary financial choices with probabilistic outcomes" (RUGGERI et al., 2020, p.2). The design of the decision problem was meticulously planned to align with the risk seeking preference for losses of medium and high probabilities, as described in CPT (KAHNEMAN and TVERSKY, 1979; TVERSKY and KAHNEMAN, 1992; KAHNEMAN, 2011; RAMOS et al., 2020). The aim was to trigger the bias alert if the decision-maker chose the alternative with risk seeking preference (alternative 2). According to the EVT, decision-makers should choose the other alternative because it has the highest EV. Kahneman and Tversky (1979; 1992) also utilized decision problems in their experiments following this typical structure in decision-making research. This approach was key to their experiments, ensuring a controlled setup that allowed the specific observation of biases in decision-making.

In step 2, the ABI approach identified decision-makers with a risk seeking preference from Step 1. If this identification occurred, the ABI tool presented a bias alert (figure 7) along with an associated explanation (figure 8). After the ABI alert, the participants were asked to answer two questions aimed at self-evaluating their awareness of a preference for seeking risk and their understanding of how their decision-making process was influenced. The first question assessed their level of agreement with the system's identification of their bias towards seeking risk. The second question gauged whether receiving an alert with an explanation about the biased preference helped them gain better insight of their decision. The participants had response options ranging from strongly agreeing to strongly disagreeing in both questions. Then, their level of awareness for risk seeking preference was measured again for comparison and evaluation purposes. Step 2 was specifically designed for participants who exhibited a risk seeking preference in Step 1, forming the focal group of our experiment.

Finally, all 154 participants answered the survey in step 3, which was the last one. Participants provided individual information, such as, gender, assessment of knowledge about decision-making theories, years of professional experience in general, years of professional experience making decisions or recommending decisions involving financial resources.

### 5.2 Participants

In this experiment, 193 participants took on the roles of employees at a company, faced with the challenge of recommending one of two options to an organizational decision-maker, both options entailing substantial financial losses. Our analysis focused on the subgroup of 101 participants who displayed risk seeking behavior in their choices and provided consistent responses. The majority of the 101 participants had many years of professional experience and were used to making decisions involving financial resources in companies with annual revenues exceeding R$ 4.8 million. Furthermore, 72% held high-level positions directly related to decision-making, such as "Superintendent / Manager", "Entrepreneur / Business Owner", and positions within executive boards, illustrating their significant influence within their organizations.

### 5.3 Formulation of statistical hypothesis

In this experiment, we wanted to assess whether our proposed ABI approach improved the level of decision-makers´ awareness for risk seeking preference. The decision-maker is the one who makes the final choice in what they perceive as the best option for the organization. For this purpose, they may consider other important elements consciously or even, unconsciously. We used the following null hypothesis (H0) and alternative hypothesis (H1):

(H0): ABI approach does not change the "level of awareness for risk seeking preference" of decision-makers.
(H1): ABI approach improves the "level of awareness for risk seeking preference" of decision-makers.

### 5.4 Determination of the experimental conditions

In this experiment, we used the following variables:

- Independent variable (IV): the use of ABI approach to identify and explain the biased risk seeking preference.
- Dependent variable (DV): level of awareness for risk seeking preference.

Before defining the level of awareness for risk seeking preference, we first define the best decision as choosing the alternative having the highest EV. We consider knowing the best decision to mean having conscious knowledge that would support the choice of one of the two alternatives. Based on Maia and McClelland (2004), for this experiment we used two possible levels of conscious knowledge of the best choice that is "the level of awareness for risk seeking preference":

- Level 0: the participant does not have any conscious knowledge specifying a preference for the best alternative.
- Level 1: the participant has conscious knowledge specifying a preference for the best alternative.

To measure the level of awareness for risk seeking preference, we asked the participant to evaluate both alternatives just after they answered the main decision problem, i.e., they made a choice by selecting one of the two alternatives. First, they evaluated alternative 1 by answering the following question: "On a scale of 0 to 10, rate how good or bad you think alternative 1 is. On the scale, 0 means it's very bad, and 10 means it's excellent." Then, we asked the same question, but to evaluate the alternative 2. Level 1 is achieved if the participant gave the highest rate to the alternative without bias ( in this experiment, it is alternative 1). If this condition is not satisfied, the participant has level 0.

## 5.5 Specifications of the procedure for assigning the subjects

In this experiment, the conditions were:

- treatment-present: Use of the ABI approach.
- treatment-absent: No use of the ABI approach.

## 5.6 Statistical Analysis

The goal of this experiment was to verify if there was a significant statistical difference between the level of awareness for risk seeking preference before and after the ABI approach alert and explanation of the risk seeking preference. This experiment involved 193 participants, of whom 154 provided consistent responses. Among these 154 valid participants: 101 of them (65.6%) chose the biased option (Risk Seeking) and 53 participants (34.4%) chose the no biased option (rational decision). The focus of this research was analyzing these 101 participants who had risk seeking preference.

Among these 101 participants, there were 84 participants (83%) who had financial decision power, i.e., who had the power to influence the decision-maker or make decisions involving financial resources. Moreover, among these 84 participants, 52% had more than 15 years of experience making decisions with financial decision power, 37% had between 5 to 15 years, and 11% had 1 to 4 years of experience.

To evaluate the impact of the ABI approach on the level of awareness for risk seeking preference among participants, we conducted the Wilcoxon signed-rank test with a significance level (alpha) of 0.05. The results provided compelling evidence, with a confidence level of at least 95%, indicating a significant difference in the participants' awareness before and after utilizing the ABI approach ($p\text{-value} < 0.05$, $p\text{-value} = 0.0039$). The p-value represents the probability of observing a difference as extreme as the one found, or even more extreme, if the null hypothesis were true. In this case, the obtained p-value of 0.0039 suggests strong evidence against the null hypothesis, suggesting that the ABI approach significantly affects awareness levels.

Furthermore, the test demonstrated a high statistical power of 0.9933, indicating a high probability of correctly detecting a significant difference when it exists. These findings underscore the substantial impact of the ABI approach in enhancing decision-makers' awareness of their risk seeking preference. Initially, all 101 participants with risk seeking preference had a level of awareness of 0. After implementing the ABI approach, 7.9% (8 participants) demonstrated an improvement in their level of awareness, achieving a score of 1.

Moreover, we found that responses indicating agreement (4) or strong agreement (5) were significantly more frequent than responses indicating disagreement (2) or strong disagreement (1) to both questions self-assessing decision making awareness after the ABI alert (table 23) and self-assessing awareness of risk seeking preference (table 24). For the self-assessed decision-making awareness, 69 participants (85.2%) agreed or strongly agreed that receiving the alert helped them gain better insight into their decision-making, while 12 participants (14.8%) disagreed or strongly disagreed (p-value: $7.905118650715595 \times 10^{-12}$). Similarly, for the self-assessed risk seeking preference, 69 participants (81.2%) agreed or strongly agreed that they have a risk seeking preference, while 16 participants (18.8%) disagreed or strongly disagreed (p-value: $1.5991900040800062 \times 10^{-13}$).

These obtained p-values indicate that the difference in proportions between the two groups is statistically significant, suggesting that the observed difference in responses is unlikely to have occurred by chance alone. We conducted a Mann-Whitney U test to compare the means of two independent samples and evaluate if they were statistically different. This nonparametric test was chosen due to the non-normal distribution of the data.

When analyzing the relationship between participants' agreement on having a risk seeking preference and their level of awareness of risk seeking preference, we found that there is no significant relationship between them. The Chi-square test yielded a p-value of 0.7063387582454752, indicating that the observed relationship is not statistically significant.

Among the 101 participants who exhibited a risk seeking preference identified by the ABI tool in the final experiment, a substantial 68% of them (69 out of the total of 101) answered that they agreed or strongly agreed with the ABI alert that they had risk seeking preference. However, a significant observation emerged as nearly 90% (62 out of 69 participants) of them remained unaware of this risk seeking inclination, even after receiving alert from the ABI tool, i.e., they remained with zero as their level of awareness of risk seeking preference. Notably, some of them justified their risk seeking choices by emphasizing their entrepreneurial mindset and the need to take risks for maximizing gains. This alignment with an entrepreneurial mindset and risk taking is in line with the principles of Effectuation theory (SARASVATHY, 2001), which suggests that entrepreneurs do not necessarily focus on maximizing expected outcomes or minimizing risks but adopt a logic of Affordable Loss (DEW & SARASVATHY, 2009).

On the other hand, 10% (7 out of 69) of the participants - identified by the ABI tool that exhibited a risk seeking preference and who answered that they agreed or strongly agreed with the ABI alert that they had risk seeking preference - enhanced their awareness of risk seeking preferences post ABI alerts. One participant aptly expressed a shift in perception. Before the ABI alert, they said that it hurts too much losing all invested money and time, but after the ABI alert they changed their perception saying that they realized that even though the loss may be heavy in the short term, they can save more in the long run and allow time to invest in another project.

While the majority of the participants remained unaware of their risk seeking preferences, a notable subset experienced a tangible shift in perception, reinforcing the complex interplay between risk awareness and decision-making contexts.

In contrast, our analysis revealed a significant relationship between participants' self-assessed decision-making awareness score after ABI alert and their level of awareness of risk seeking preference. The Chi-square test resulted in a p-value of 0.000192753383318726777 with 4 degrees of freedom, indicating a strong association between these variables. Further examination reveals that participants who strongly agreed with the idea that receiving the ABI approach alert helped them gain better insight into their decision-making were more likely to experience an improvement in their risk seeking awareness level from 0 to 1. Out of the 12 participants who strongly agreed, 5 participants showed this improvement. This finding emphasizes the importance of decision-makers' understanding of their decision processes in recognizing and adjusting for their own risk biases.

### 5.7 Preliminary Conclusion

The ABI approach was the answer to the main RQ of this study. Findings from the final experiment provided evidence indicating that the use of the ABI approach may contribute to organizational decision-makers to identify that they are subjected to risk seeking preference during the decision-making process. Therefore, this was an evidence that an automatized method over a well-founded ontology of intuitive decision-making allows to support organizational decision-makers to identify that they are subjected to individual risk seeking preference during the GO/KILL decision-making. Our ontology of intuitive decision-making under risk and uncertainty allowed us to precisely characterize and identify the decision-makers' preference for risk seeking and it was used by the ABI tool to alert the decision-makers how their risk preferences may be biased and how this situation may occur.

## 6 CONCLUSION

This section presents the summary, results, contributions, limitations, and future work of this research.

### 6.1 Research Summary

This research explored the critical area of decision-making within organizational contexts, focusing on the challenge of an individual organizational decision-maker identifying during decision-making that they are influenced by risk seeking preference, which may lead them to reject the favorable alternative and to choose the riskier one. This research focused on individual GO/KILL decisions under risk, associated with an organization's goal, with no time pressure and involving high losses of medium and high probability.

Addressing the main research question of "How to support an organizational decision-maker to identify that they are subjected to individual risk seeking preference during a GO/KILL decision-making?", this research introduced the Automatic Bias Identification (ABI) approach. This approach was supported by the Intuitive Decision-Making Ontology, developed specifically for this research. The ontology provided a framework for understanding the contexts and conditions under which risk preferences were likely to manifest in intuitive decision-making under risk and uncertainty. The proposed ontology was based on the Cumulative Prospect theory, mainly on the fourfold pattern of risk attitudes from (TVERSKY and KAHNEMAN, 1992).

The ABI approach leveraged this ontology to automatically identify risk seeking preferences in decision-making processes under risk and elucidate the context and biases influencing these preferences. The application of this approach was demonstrated through the ABI tool, which is a computational tool that uses the proposed ontology to automatically identify and explain the risk seeking preference. The ABI approach was assessed by means of an experiment.

### 6.2 Results

The ABI approach was evaluated in a final experiment, addressing the main research question. In this experiment, 193 participants took on the roles of employees at a company, faced with the challenge of recommending one of two options to an organizational decision-maker, both options entailing substantial financial losses. Our analysis focused on the subgroup of 101 participants who displayed risk seeking behavior in their choices and provided consistent responses. The majority of the 101 participants had many years of professional experience and were used to making decisions involving financial resources in companies with annual revenues exceeding R$ 4.8 million. Furthermore, 72% held high-level positions directly related to decision-making, such as "Superintendent/Manager", "Entrepreneur/Business Owner", and positions within executive boards, illustrating their significant influence within their organizations.

This final experiment captured the essence of CPT, emphasizing the theory's prediction that decision-makers have a higher propensity for risk seeking in the domain of medium to high probabilities of losses. These findings underscore the predictive power of CPT in the domain of loss and the importance of considering its implications in decision-making research.

The evaluation of the final experiment provided evidence indicating that the use of the ABI approach, which encompasses an automatized method over a well-founded ontology of intuitive decision-making, contributed to organizational decision-makers to identify that they are subjected to individual risk seeking preference during the GO/KILL decision-making. Our ontology of intuitive decision-making under risk and uncertainty allowed to precisely characterize and identify the decision-makers' preference for risk seeking and it was used by the ABI tool to alert decision-makers about a situation in which their risk preferences may be biased and how this situation may occur. Therefore, the ABI approach contributed to answering the main research question.

## 6.3 Contributions

The scientific contribution of this research was to improve the understanding by decision-makers of their biased risk preferences in organizational decision-making under risk.

We contributed to the area of Information System by developing and evaluating the ABI Approach, including the intuitive decision-making ontology for decision-making under risk and uncertainty. The ontology provides a contextualized explanation of the bias for the specific situation and therefore constitutes a new opportunity of application in the Information System. The ontology was codified into a logical data schema to serve as a knowledge base accessed by the ABI tool that allowed the automatic identification of risk seeking preference and the explanation of this preference contextualized for each scenario. The ABI tool is a computational solution and is the pilot implementation of the ABI approach.

In addition, the ABI approach and the intuitive decision-making ontology contributed to the mission of the 2020's for decision support systems (DSS or digital coaching systems) that is to integrate human intelligence with advanced system capabilities for digitalization (CARLSSON and WALDEN, 2019). "In order to make it work, human systems users need context relevant advice (in real time, with real data and information) that is adapted to their cognitive abilities and background knowledge (i.e., advice they can understand and use)" (CARLSSON and WALDEN, 2019, p. 240).

This research also contributed to the integration of Information System and Behavioral Economics, as we extended the application of Cumulative Prospect theory from Behavioral Economics in a novel way by automating the identification and explanation of risk seeking preferences. The ABI approach transforms complex theoretical insights into actionable, real-time guidance within organizational decision-making contexts. This approach democratizes access to sophisticated behavioral economics insights by eliminating the need for specialized personnel to identify and explain these preferences. By automatically translating Cumulative Prospect theory concepts into business language that is contextualized to the decision-making situation, the ABI approach makes enhanced decision support more accessible and applicable to a broader range of organizations and decision-makers. This includes those who may never have had the opportunity to learn about CPT, as well as organizations that cannot afford the training or hiring of a decision observer, thereby facilitating widespread adoption and enhancing decision-making processes with deep behavioral insights.

Another significant contribution of the ABI approach is its ability to automatically collect historical data on risk seeking preferences and decision outcomes within organizations. This dataset allows for analysis to identify decision-making patterns and biases, offering insights that can enhance strategic management and organizational performance. Additionally, the data provides researchers with unique opportunities to study the effects of bias awareness on decision-making (SAYÃO and BAIÃO, 2023).

Considering the educational contributions of this research, the insights and data generated by the ABI tool can be used for training and educational purposes, helping decision-makers understand and mitigate their biases more effectively. This could be particularly useful in academic settings or corporate training programs focused on decision-making and risk management.

## 6.4 Limitations and Future Work

In this research, Expected Value theory was utilized as the standard for rational decision-making. However, EVT may not capture the full complexity of decisions within organizational contexts, such as those involving strategic considerations such as long-term growth and market positioning. Similarly, our final experiment revealed limitations in using Cumulative Prospect theory as the sole perspective through which risk seeking behavior is identified by the ABI approach.

To address these limitations, future enhancements to the ABI approach could involve integrating other theories like Effectuation (SARASVATHY, 2001) and Affordable Loss (DEW and SARASVATHY, 2009)(MARTINA, 2020), which offer a broader perspective on decision-making. These theories accommodate the various objectives and risk capacities of decision-makers, particularly those with an entrepreneurial mindset, who may prioritize different outcomes than what EVT and CPT typically measure.

Other limitations of this research:

- The proposed ABI approach, designed with inherent flexibility, currently specializes in identifying risk seeking preference within domains involving losses of medium to high probability. The intuitive decision-making ontology underpinning it recognizes the four risk preferences outlined in the fourfold pattern of risk attitudes of CPT. Future iterations expanding the ABI tool to cover the full spectrum of risk preferences, as suggested by the ontology, would greatly enhance its comprehensiveness and effectiveness in diverse decision-making contexts.
- In this research, we focus on decision scenarios that present only two alternatives, characterized solely by their financial values and probabilities of success as the primary criteria. This model is reflective of a common approach in decision-making research, which Ruggeri et al. (2020, p.2) describe as "binary financial choices with probabilistic outcomes", a structure that facilitates controlled and structured analysis. While this model is useful for understanding the influence of cognitive biases on decision-making, it also simplifies the complexity inherent in organizational decisions, which may include additional factors.
- This research assumes that the probabilities of each decision alternative are estimated before using the ABI approach, supported by research and practical methods that facilitate these estimations (LOVALLO et al., 2020; MAIZLISH and HANDLER, 2005). One participant of the final experiment expressed that his company once hired 3 different lawyers to get their risk estimates on pursuing a specific path and the CEO made the final decision based on these estimates.

Future work should focus on enhancing the ABI approach to cover a wider array of decision-making criteria. This includes evaluating stakeholder impacts, resource availability, and the effects of decisions on ongoing projects, as well as assessing the organization's competitive advantage. Additionally, future enhancements should consider factors intrinsic to decision-makers, such as risk appetite and years of financial decision-making experience, alongside the unique demands of their industry. Such expansion will yield a richer, more holistic perspective on organizational decision-making processes.

Future work for further development of the ABI approach could focus on the following areas:

- Training Tool Development: incorporate the ABI approach into a training mechanism for decision-makers to identify and address their risk seeking preferences during the decision-making process. By integrating this approach into training programs, decision-makers can enhance their self-awareness and develop strategies to mitigate the influence of biases, resulting in more informed and balanced decision-making. This training can be particularly effective for new decision-makers without financial resource power and decision-makers involved in financial resource decisions for less than five years, regardless of their professional experience. As suggested by our final experiment analysis, decision-makers in these specific groups were more likely to increase their level of awareness about their biased risk seeking preferences.
- Expansion of Cognitive Bias and Decision-Making Theories Exploration: enhance the ABI approach by integrating additional cognitive biases and decision-making theories relevant to organizational contexts. These adaptations aim to improve decision-making processes and outcomes in organizational settings, making the ABI tool more versatile and applicable to a range of business scenarios.
- Artificial Intelligence Integration: explore the use of Artificial Intelligence to more accurately identify and analyze decision-makers' risk preferences and cognitive biases. Leveraging AI could offer more sophisticated, data-driven insights and improve the predictive capabilities of the ABI approach. This aspect is already being explored by researchers involved in the "Support for Decision-Making in Knowledge-Intensive Scenarios Addressing Behavioral and Cognitive Perspectives through Behavioral Economics" project, such as (SAYÃO and BAIÃO, 2023).

- Feedback Mechanisms: integrate feedback loops within the ABI approach to provide continuous learning opportunities from past decisions. This aspect is already being addressed by the researchers involved in the "Support for Decision-Making in Knowledge-Intensive Scenarios Addressing Behavioral and Cognitive Perspectives through Behavioral Economics" projectB. y integrating feedback loops, decision-makers can learn from their past decisions and adapt their approach accordingly, fostering continuous improvement.
- Longitudinal Study: conduct a long-term study to observe the impact of the ABI approach supporting an organizational debiasing process on reducing the decision-maker's bias over time. This would involve observing decision-makers over an extended period and analyzing the long-term effects to assess whether the ABI approach leads to behavioral changes and if it is beneficial for the company. This comprehensive longitudinal study would provide valuable insights into how the ABI approach influences decision-making patterns and performance over time.

As future work, we make several suggestions that can be combined or considered individually, including:

(i) Considering the other three risk attitude patterns addressed by the CPT (KAHNEMAN, 2011): risk aversion for gains of medium and high probability; risk seeking for gains and risk aversion for losses of low probability;

(ii) Considering group GO/KILL decisions in addition to individual GO/KILL decisions;

(iii) Considering other criteria besides financial values and probability of success, such as impact on stakeholders, availability of resources (e.g., budget, staff, equipment), impact on other ongoing projects, technical criteria (e.g., technical feasibility, complexity, and potential risks associated with the project) and potential impact on the organization's competitive advantage;

(iv) Considering decision problems with more than two alternatives;

(v) Considering other cognitive biases and other theories of BE, such as Nudge.